\documentclass{jhep3}
\usepackage{amssymb, amsmath}
\usepackage{graphicx}
\usepackage{cite}
\usepackage{empheq}

\newcommand{\cH}{\mathcal{H}}

\def\be{\begin{equation}}
\def\ee{\end{equation}}
\def\bea{\begin{eqnarray}}
\def\eea{\end{eqnarray}}

\title{ \bf{Constructing near-horizon geometries in supergravities with hidden symmetry}}

\author{Hari K. Kunduri$^a$\footnote{hkunduri@phys.ualberta.ca } \  and James Lucietti$^b$\footnote{j.lucietti@ed.ac.uk } \\ \\
\small \sl $^a$ Theoretical Physics Institute,
Department of Physics, University of Alberta, \\ \small \sl Edmonton, AB, T6G 2J1, Canada
\\ \\ \small \sl $^b$ School of Mathematics and Maxwell Institute of Mathematical Sciences, University of Edinburgh,  King's Buildings, Edinburgh, EH9 3JZ, UK }

\date{}

\abstract{
We consider the classification of near-horizon geometries in a general two-derivative theory of gravity coupled to abelian gauge fields and uncharged scalars in four and five dimensions, with one and two commuting rotational symmetries respectively. Assuming that the theory of gravity reduces to a 3d non-linear sigma model (as is typically the case for ungauged supergravities),  we show that the functional form of any such near-horizon geometry may be determined. As an example we apply this to five dimensional minimal supergravity. We also construct  an example of a five parameter near-horizon geometry solution to this theory with $S^1 \times S^2$ horizon topology. We discuss its relation to the near-horizon geometries of the yet to be constructed extremal black rings with both electric and dipole charges.}

\begin{document}

\section{Introduction}
The first microscopic derivation of the Bekenstein-Hawking black hole entropy was for certain {\it five}-dimensional supersymmetric black holes in string theory\cite{SV}. The simplest such black hole is the extremal Reissner-N\"ordstrom (RN) solution, which is parameterised by just its electric charge $Q$. A simple setup for embedding such black holes in string theory is as the dimensional reduction of a D1-D5-P intersecting brane solution to type IIB supergravity on $T^5$, where $Q=N_1=N_5=P$~\cite{CM}.  These entropy calculations were quickly generalised to a rotating generalisation of the five dimensional RN solution that preserves supersymmetry, the BMPV black hole~\cite{BMPV}, which is parameterised by its electric charge $Q$ and angular momentum $J_1=J_2$.  In fact, more generally there exists a consistent truncation\footnote{An explicit Kaluza-Klein reduction for this is given in \cite{Her}.} of IIB supergravity on $T^5$ to five dimensional minimal supergravity -- the bosonic sector of this theory is simply Einstein-Maxwell coupled to a Chern-Simons term -- so any solution to this five dimensional theory may be embedded into type IIB supergravity. Subsequently, successful entropy calculations were performed for four-dimensional supersymmetric non-rotating black holes in $\mathcal{N}=4$ and $\mathcal{N}=8$ supergravity \cite{MS1}, which arise upon reduction of IIA supergravity on $K3 \times S^1 \times S^1$ and $T^4 \times S^1 \times S^1$ respectively.

A key assumption of these early string theory calculations is that a stationary black hole is uniquely  specified by the appropriate conserved charges $(M,J_1,J_2,Q)$. This assumption was inspired by the classic four dimensional black hole uniqueness theorems which indeed state that there is only one black hole solution given $(M,J,Q)$ (the Kerr-Newman solution). It has turned out that this remarkable property of four dimensional black holes does not generalise to five dimensions. This was revealed by the remarkable discovery of the black ring, an asymptotically flat vacuum black hole solution with $S^1\times S^2$ (spatial) horizon topology~\cite{ER}. In conjunction with the Myers-Perry black hole, this explicitly demonstrates that five dimensional black holes are not uniquely specified by their conserved charges (for a range of $M$ and $J_i$ there are two black rings and one Myers-Perry black hole).

Subsequently, generalisations \cite{emparan,EEF,EEMR} of the black ring solutions were found in Einstein-Maxwell-CS theory with coupling corresponding to the bosonic sector of minimal supergravity. A new feature which arises is that such solutions can carry a non-conserved magnetic dipole charge\footnote{In contrast to four dimensional Einstein-Maxwell theory, there is no magnetic conserved charge in five dimensions. The dipole charge is the closest analogue, but can only be defined for black ring horizon topology.} $\mathcal{D}$. This leads to a continuous non-uniqueness for such black holes, on top of the discrete non-uniqueness of the vacuum solutions~\cite{emparan}. Thus stationary black hole solutions to five dimensional Einstein-Maxwell-CS may carry all such charges $(M,J_1,J_2, Q,\mathcal{D})$. In fact, although not a conserved charge, the dipole charge still appears in the first law of black hole mechanics~\cite{emparan,CH}
\be
\delta M = \frac{\kappa \delta A_H}{8\pi} + \Omega_i \delta J_i + \Phi_H \delta Q + \Psi_H \delta \mathcal{D} \; ,
\ee
indicating that it must play a fundamental role in the quantum description of such objects.

The additional feature of dipole charge makes it more difficult to map out the space of black hole solutions in five-dimensional supergravity theories.  One may straightforwardly add global electric charges to vacuum solutions by performing duality transformations. While this technique is successful when applied to black holes with spherical topology, singularities inevitably arise unless the initial black rings already carry dipole charge~\cite{EEF}. A strategy to circumvent this problem is to dimensionally reduce to two dimensions (we restrict to stationary solutions with $\mathbb{R} \times U(1)^2$ isometry) and use integrability methods to generate new solutions\cite{Fig1}. This approach has proved extremely successful in producing novel vacuum solutions such as the doubly-spinning ring \cite{PS} and the Black Saturn \cite{BSat}, and is capable of dealing with dipole charge.

In this work we will pursue an alternative strategy by focussing on \emph{extremal} black holes.  An extremal black hole possesses a unique near-horizon geometry, a solution in its own right which gives a precise description of spacetime near the event horizon. While technically simpler to analyse, near-horizon geometries retain properties of the parent black hole that are intrinsic to the horizon, in particular the horizon topology and geometry. Given a near-horizon geometry with no reference to how it connects to a full black hole solution, it is of interest to calculate the conserved charges.  It turns out for black holes with spherical topology, it is possible to directly calculate $(Q,J_1,J_2)$ from the near-horizon geometry alone. For black rings the ambiguity associated with identifying the $S^1$ direction at spatial infinity in the near-horizon means one cannot read off the angular momentum uniquely in this direction, however, one can compute the angular momentum associated to the $S^2$, the electric charge and the dipole charge (see e.g. \cite{HOT}).

Classifications of near-horizon geometries of rotating extremal black holes have been achieved in four-dimensional $\mathcal{N}=2$ supergravity\cite{LP,KL4d} (Einstein-Maxwell theory with a non-positive cosmological constant). In five dimensions, a full classification was made for supersymmetric near-horizon geometries in $\mathcal{N}=1$ ungauged supergravity \cite{Reall}, and in $\mathcal{N}=1$ gauged supergravity under the assumption of \emph{two} commuting rotational symmetries \cite{KLR:2006, KL:2007}. In higher than five dimensions a variety of classification results for supersymmetric near-horizon geometries have been obtained~\cite{GMR, GP}. In the non-supersymmetric case, much less is known;  progress has been restricted to the pure vacuum \cite{KLvac} and static near-horizon geometries \cite{KLstatic}.  Recently, the classification of vacuum near-horizon geometries in $D$ dimensions with $U(1)^{D-3}$ spatial isometries was achieved \cite{HI} by exploiting the fact the field equations are equivalent to a three dimensional $SL(D-2,\mathbb{R})/SO(D-2)$ non-linear sigma model.  This extra structure is enough to allow for a direct integration of the near-horizon data. One of the aims of this paper is apply this method to more general theories with matter.

In this paper, we consider a general second order theory of gravity coupled to uncharged scalars and Abelian gauge fields in $D=4,5$. Focussing on extremal black holes solutions with $D-3$ commuting rotational symmetries, we show that if the theory, when reduced to three dimensions, can be cast as a non-linear sigma model with a symmetric space as the target space, then it is possible to integrate directly to determine the explicit functional form of all near-horizon geometries. Our main motivation for investigating this problem was to apply it to supergravity theories which arise from truncations of supergravities in 10 and 11 dimensions. For example, we will apply our general results to $D=5$ minimal ungauged supergravity, which has such a non-linear sigma model description with a target coset $G_{2,2}/SO(4)$ \cite{CJLP}.  It should be emphasised though that our results are valid in a much wider context and could even be applied to theories such as $D=4$, $\mathcal{N}=8$ supergravity, which when reduced to three dimensions is a non-linear sigma model with target space $E_{8,8}/SO(16)$  \cite{CJ:1979}. However, in practice, the main obstacle to completing a full classification is that there are a large number of non-linear algebraic constraints between the parameters in the general solution.

This paper is organized as follows. In section \ref{general}, we review the construction of using symmetries to rewrite solutions in terms of scalars. In section \ref{sec:NHG}, we show how these scalars are related to the canonical form of near-horizon geometries established in \cite{KLR}, and give gauge invariant formulas for the charges. We also show how one can integrate for the general solution in the case the theory can be cast as a non-linear sigma model. In section \ref{sec:5dsugra} we consider applications of this method to minimal supergravity theory in $D= 5$ and also construct a five parameter family of near-horizon geometries with $S^1\times S^2$ horizon topology. Finally, in section \ref{discussion} we discuss our results and make some comments regarding the space of extremal black rings in minimal supergravity.

\section{Theories of gravity with hidden symmetry}\label{general}
Consider a general 2-derivative theory describing Einstein gravity coupled to abelian gauge fields $A^I$ ($I=1\ldots N$) and uncharged scalars $\phi^A$ ($A=1 \ldots M$) in $D=4,5$ dimensions, with action
\be
\label{gentheory}
 S = \int d^D x \sqrt{-g} \left(R -  f_{AB}(\phi)  d\phi^A \cdot d \phi^B - V(\phi) - g_{IJ}(\phi) F^I \cdot F^{J} \right) + S_{{\rm top}},
\ee
where $F^I \equiv dA^I$, $V(\phi)$ is an arbitrary scalar potential, and
\be
 S_{{\rm top}} = \begin{cases} \frac{1}{2} \int h_{IJ}(\phi) F^I \wedge F^J & \text{if } D=4  \\
 -\frac{1}{6} \int C_{IJK}F^I \wedge F^J \wedge A^K & \text{if } D=5 \; , \end{cases}
\ee
where $C_{IJK}$ are constants.

We begin our discussion by considering general $U(1)^{D-3}$-invariant solutions. One can always decompose $U(1)^{D-3}$-invariant metrics as
\be
g= \gamma_{ij}(dx^i+\omega^i)(dx^j+\omega^j) +\gamma^{-1} h_{\mu\nu} dx^\mu dx^\nu
\ee
where $i,j=1, \dots D-3$ and $m_i=\partial/ \partial x^i$ are the Killing vector fields. We introduce the 3d base space $\mathcal{M}_3$ with coordinates $x^{\mu}$ and a Lorentzian metric $h_{\mu\nu}$. The quantities $\omega^i=\omega^i_\mu dx^\mu$ are 1-forms on $\mathcal{M}_3$ and $\gamma_{ij}$ are functions on $\mathcal{M}_3$ and we write $\gamma= \det \gamma_{ij}$.  The base space $\mathcal{M}_3$ is in fact the orbit space of the $U(1)^{D-3}$ isometry, and we can think of the spacetime as a $T^{D-3}$ fibration over $\mathcal{M}_3$.

 We also assume that the spacetime Maxwell fields $F^I$ and scalar fields $\phi^A$ are invariant under the Killing fields $m_i$. It follows that $\phi^A$ are functions on the 3d base space $\mathcal{M}_3$. The Maxwell fields may be written in terms of a set of potentials defined on $\mathcal{M}_3$, as we now show.

Define the magnetic 1-forms $B^I_i = - i_{m_i} F^I$, which are closed as a consequence of $dF^I=0$ and the assumed invariance of the matter fields under $m_i$. Hence we can introduce the magnetic potentials $b^I_i$ via
\be\label{magnetic}
B^I_i =- i_{m_i} F^I = db_i^I  \; .
\ee
Notice that $\mathcal{L}_{m_i} b^I_j = -i_{m_i} i_{m_j} F^I$ is a constant by standard arguments, and thus if one $m_i$ vanishes somewhere (as will be the case for us), we deduce that $\mathcal{L}_{m_i} b_j^I=0$, i.e. $b_i^I$ are functions on $\mathcal{M}_3$.
Similarly we may define the electric 1-forms
\begin{equation}\label{electric}
E_I = -  g_{IJ}(\phi)  \; i_{m_1}  \cdots i_{m_{D-3}} \star F^J
\end{equation} which satisfy
\begin{equation}
dE_I = -(-1)^{D-3} i_{m_1}\cdots i_{m_{D-3}}  dS_I \; ,
\end{equation} where $S_I$ is the topological term arising in the Maxwell equation for $F^I$
\bea
d ( g_{IJ}(\phi) \star F^J - S_I )=0  \; , \quad
 S_I =
  \begin{cases}
  \frac{1}{4}h_{IJ}(\phi)F^J & \text{if } D =4 \\
  - \frac{1}{8}C_{IJK}A^J \wedge F^K       & \text{if } D=5  \; .
  \end{cases}
\eea Explicitly, we have
\begin{equation}
dE_I = d s_I \;, \quad s_I =   \begin{cases}
  \frac{1}{4}h_{IJ} db^J & \text{if } D =4 \\
   \frac{1}{8}C_{IJK}\left( b_1^J db_2^K - b_2^J db_1^K \right)      & \text{if } D=5  \; .
  \end{cases}
\end{equation} Hence there exists a scalar potential $\mu_I$ given by\footnote{In $D=5$, although $s_I$ appears to break covariance with respect to the two Killing fields $m_i$, it is invariant under the $SL(2,\mathbb{Z})$ symmetry which acts on the $m_i$.}
\begin{equation}\label{mudef}
d\mu_I \equiv E_I - s_I \; .
\end{equation}
It is clear that $\mathcal{L}_{m_i} \mu_I=0$ and thus $\mu_I$ are also functions on $\mathcal{M}_3$.  It is worth emphasising that given the potentials $(b^I_i, \mu_I)$ one can completely reconstruct the Maxwell fields $F^I$ using the identity
\begin{equation}
F^I =\frac{1}{\gamma} \left[ g^{IJ}(\phi)\star (m_{D-3} \wedge \cdots m_1 \wedge E_J) - \gamma\;  \gamma^{ij} m_i \wedge B_j^I \right]
\end{equation} where $g^{IJ}(\phi)$ is the inverse of $g_{IJ}(\phi)$, which can be verified directly from the definitions (\ref{magnetic}) and (\ref{electric}).

Finally, we introduce the twist potentials. First define the twist one-forms
\be
\Omega_i = \star \, ( m_1 \wedge \cdots m_{D-3} \wedge dm_i )
\ee
which satisfy $d\Omega_i = 2(-1)^{D-3} \star ( m_1 \wedge \cdots m_{D-3} \wedge \text{Ric}(m_i) )$, where we have defined the 1-form $\text{Ric}(m_i) \equiv \text{Ric}(m_i, \cdot)$ and $\text{Ric}$ is the Ricci tensor of the spacetime metric $g$. Using the Einstein equation one can show that
\be
d\Omega_i = 4E_I \wedge B_i^I =  \begin{cases} -2d\left( b^I d\mu_I - \mu_I db^I\right) & \text{if } D=4 \\ -d\left[ b_i^I \left( 4 d\mu_I + \frac{1}{6}C_{IJK}( b_1^J db_2^K - b_2^J db_1^K )\right)\right] & \text{if } D=5  \end{cases}
\ee
which implies the existence of the twist potentials $Y_i$ given by
\be
\label{Ydef}
 dY_i = \Omega_i + \begin{cases} 2\left( b^I d\mu_I -   \mu_I db^I\right) & \text{if } D=4 \\  b_i^I \left( 4d\mu_I + \frac{1}{6}C_{IJK}( b_1^J db_2^K - b_2^J db_1^K )\right) & \text{if } D=5  \; . \end{cases}
\ee
It  is clear that $\mathcal{L}_{m_i} Y_j=0$ and thus $Y_i$ are also functions purely on $\mathcal{M}_3$.

The potentials $(b_i^I, \mu_I, Y_i)$ we have introduced are only defined up to certain gauge transformations.  For completeness we give these:
\bea
\label{gauget1}
&&b_i^I \to b_i^I + c_i^I  \\  \label{gauget2}
&&\mu_I \to  \mu_I + \begin{cases} \mu_I^0  & \text{if } D=4 \\ \mu_I^0 -\frac{1}{8} C_{IJK} ( c_1^J b_2^K-c_2^J b_1^K)& \text{if } D=5 \end{cases}  \\ \label{gauget3}
&& Y_i \to Y_i +\begin{cases} y_i + 2(c^I\mu_I - \mu^0_I b^I) & \text{if } D=4 \\ y_i  +4c_i^I \mu_I +\frac{1}{6} C_{IJK} ( c_2^I b_1^J -c_1^I b_2^J)( b_i^K+ 2c_i^K) & \text{if } D=5 \end{cases}
\eea
where $c_i, \mu_I^0, y_i$ are constants.  The other scalars $\gamma_{ij}$ are clearly gauge invariant.

Before moving on, it is worth noting that if we are on an ``axis" of rotational symmetry, i.e. where some linear combination $v^i m_i$ vanishes,  then
\be
v^iB^I_i=0 \qquad E_I=0 \qquad \Omega_i=0 \; .
\ee
The first relation implies $v^i b^I_i$ are constant on such axes. In 4d it is thus clear that all potentials $b^I$, $\mu_I$ and $Y$ are constant on such an axis. However, in 5d any other linear combination of the $b^I_i$ need not be a constant on such axes, and furthermore due to the CS terms in (\ref{mudef}) and (\ref{Ydef}) the other potentials $\mu_I, Y_i$ need not be constant either (although later in the context of near-horizon geometries we will see that all potentials are constant on the axes).

So far we have shown that we can represent $U(1)^{D-3}$-invariant solutions $(g,F^I, \phi^A)$ to the above theory, by a 3d metric $h_{\mu\nu}$ and a set of scalars/potentials $\Phi^M=( \gamma_{ij}, Y_i, b_i^I, \mu_I, \phi^A)$, which are all defined on the 3d space $\mathcal{M}_3$.   Indeed, generically such solutions may be equivalently derived from a 3d theory of gravity coupled to a harmonic map
\be
S[h,\Phi]=\int_{\mathcal{M}_3}  \sqrt{-h} \left[ R^{(3)}-  h^{\mu\nu} G_{MN}(\Phi) \partial_\mu \Phi^M \partial_\nu \Phi^N \right]
\ee
where $R^{(3)}$ is the Ricci scalar of $h$ and $G_{MN}(\Phi)$ may be thought of as a metric on the scalar manifold with coordinates $\Phi^M$.  We will assume that our scalar manifold is a symmetric space $G/H$ with the bi-variant metric
\be
G_{MN}(\Phi) d\Phi^M d\Phi^N = \frac{1}{4m} \textrm{Tr}( M^{-1} dM \; M^{-1} dM)
\ee
where $M$ is a matrix constructed from the set of scalars $\Phi^M$ and $m$ is a normalisation constant chosen for later convenience. We assume the matrix $M$ is an $n \times n$ hermitian (or symmetric if $M$ is real), positive, unimodular, representative of $G/H$.  This is sufficient to capture many supergravities of interest in four and five (and higher) dimensions.
In other words, our 3d theory of gravity is coupled to a non-linear sigma-model. The equations of motion for this 3d sigma model are
\bea
\label{Richeq}
&&R^{(3)}_{\mu\nu} = \frac{1}{4m} \textrm{Tr}\,  \left ( M^{-1} \partial_\mu M \, M^{-1} \partial_\nu M\right)  \\
\label{Meq}
&&D^{\mu} ( M^{-1}\partial_{\mu} M )=0
\eea
where $D$ is the connection associated to the base metric $h$.

In the absence of the matter fields this must reduce to the vacuum $SL(D-2,R)$ nonlinear sigma-model~\cite{Maison}. In this case $m=1$ and $M$ is given by  the real symmetric $(D-2) \times (D-2)$ matrix
\be
\label{chi}
\chi= \left( \begin{array}{cc} \gamma_{ij}+ \gamma^{-1}Y_iY_j & -\gamma^{-1}Y_i \\ -\gamma^{-1}Y_j &  \gamma^{-1} \end{array} \right)   \; .
\ee
With the matter fields, $M$ is an $n \times n$ matrix, which in the vacuum limit  must be equivalent to a block diagonal form with $m$ copies of the $SL(D-2,R)$ matrix $\chi$ and an identity matrix, such that $n \geq m(D-2)$. The explicit form of $M$ will depend on the precise theory we are working with.

\section{Near-horizon geometries in theories with hidden symmetry}
\label{sec:NHG}
\subsection{Determining the geometries}
Now consider any extremal black hole solution to the theory (2.1), possessing $D-3$ rotational Killing fields $m_i$ which leave the matter invariant, with compact horizon sections $\mathcal{H}$ of non-toroidal topology. As shown in~\cite{KLR}, the near-horizon geometry is generically a non-trivial fibration of the horizon over AdS$_2$. Explicitly, it may be written as
\bea
&&g_{NH} = \Gamma(x) \left[ -\frac{{r}^2 dv^2}{\ell^2} +2dvd {r} \right] +L^2 \left[ \frac{dx^2}{ \gamma(x)} + \gamma_{ij}(x) \left( d\phi^i+\frac{k^i {r} dv}{L^2} \right) \left(d\phi^j +\frac{k^j {r} dv}{L^2} \right) \right]  \label{so21form} \nonumber  \\
 &&F^I_{NH} = d \left [ \frac{e^I \, {r}dv}{L} + L b^I_i(x) \left( d\phi^i + \frac{k^i {r}dv}{L^2}  \right) \right]   \nonumber \\
 &&\phi^A_{NH} = \phi^A(x)   \label{canonical}
\eea
where $\gamma =\det \gamma_{ij}$,  $r=0$ is the horizon, $\partial / \partial v$ is Killing field normal to the horizon, $(x,\phi^i)$ are coordinates on sections of the horizon $\mathcal{H}$, and $m_i = L^{-1} \partial / \partial \phi^i$. The coordinate $x$ is defined purely geometrically by $dx= -L^{-1} \, i_{{m_1}} \cdots i_{m_{D-3}} \epsilon_{D-2}$, where $\epsilon_{D-2}$ is the volume form of the metric on sections of the horizon, as in~\cite{KLvac}.\footnote{If the horizon orbit space $\mathcal{H}/U(1)^{D-3}$ is simply connected, this 1-form is globally exact. Therefore the function $x$ always exists for non-toroidal horizon topology, since the orbit space is then a closed interval. This function may then be used as a coordinate everywhere except where some linear combination of $m_i$ vanish (which only occurs at the endpoints of this interval).} The coordinate ranges are $\phi^i \sim \phi^i +2\pi$ and $-x_0 \leq x \leq x_0$ with $\gamma(\pm x_0)=0$. For definiteness, we will define our spacetime orientation by $dv \wedge dr \wedge dx \wedge d\phi^1 \wedge d\phi^2 >0$. The quantities $k^i, e^I$ are constants, and $\ell, L$ are two length scales associated to AdS$_2$ and $\cH$ respectively, introduced for later  convenience. This form reveals an enhanced $SO(2,1)\times U(1)^{D-3}$ symmetry whose orbits are generically $T^{D-3}$ fibrations over AdS$_2$. Note there is a scaling freedom
\bea
\label{scaling1}
  (\ell, L,x,v,\Gamma, e^I, b^I_i) \to (s\ell, sL, \; s^{-1}x, \; s^2 v,  \; s^{-2}\Gamma, \; s^{-1}e^I, \; s^{-1}b^I_i)
\eea
where $s$ is a constant and will be used later to simplify the solution.

We are interested in the explicit classification of such geometries. One of the aims of this note is to show that in a subset of theories of the form (\ref{gentheory}) it is easy to determine the functional form of such geometries. The subset is such that the KK reduction on $T^{D-3}$ yields 3d gravity coupled to a non-linear sigma model as discussed in the previous section.

To identify the 3d data associated to the class of near-horizon geometries above we first need to identify the $U(1)^{D-3}$ isometries. This can be done at the level of coordinates by setting  $x^i =L \phi^i$, so $x^{\mu}= (v,r,x)$ are our coordinates on $\mathcal{M}_3$.
Now, we can read off the 3d data associated to the metric, which consists of the base metric $h_{\mu\nu}$, 1-forms $\omega^i$ and scalars $\gamma_{ij}$:
\bea
&&h_{\mu\nu} dx^\mu dx^\nu =  L^2 dx^2 + Q(x)[ -\ell^{-2} {r}^2 dv^2 +2dvd{r} ]  \\
&&\omega^i = \frac{k^i {r}dv}{L}  \\
&&\gamma_{ij}=\gamma_{ij}(x)
\eea
where  we have defined the function
\be
\label{Qdef}
Q(x) \equiv \Gamma(x)\gamma(x)  \; .
\ee
The 3d data associated to the Maxwell fields may be determined by
\bea
b_i^I  &=& b_i^I(x)    \\
E_I &=& g_{IJ}(\phi) L \Delta^J(x) dx
\eea where
\begin{equation}
\Delta^I\equiv \frac{e^I + b_i^I k^i}{L \Gamma}.
\end{equation}
Therefore, from (\ref{mudef}) we deduce that $\mu_I$ are functions of $x$ only,  and that
\bea
\label{NHmu}
\mu'_I &=& g_{IJ}(\phi) L\Delta^J -  \begin{cases}
  \frac{1}{4} h_{IJ} {b^J}' & \text{if } D =4 \\
   \frac{1}{8}C_{IJK}\left( b_1^J {b_2^K}' - b_2^J {b_1^K}' \right)      & \text{if } D=5  \; .
  \end{cases}
\eea  It is straightforward to check
\begin{equation}
\Omega_i = \frac{\gamma_{ij}k^j}{\Gamma} dx
\end{equation} and hence from (\ref{Ydef}) it follows the twist potentials are functions of $x$ alone and can be determined from
\be
Y_i'  = \frac{\gamma_{ij} k^i}{\Gamma} + \begin{cases}  2\left( b^I \mu_I' - \mu_I b^{I'}\right) & \text{if } D=4  \\ b_i^I  \left[4 \mu_I' + \frac{1}{6}C_{IJK}( b_1^J {b_2^K}' - b_2^J {b_1^K}' )\right]  & \text{if } D=5 \; .  \end{cases}
\ee

Therefore all scalars/potentials depend only on $x$. Hence the metric on the scalar manifold, and hence the matrix $M$, only depend on $x$, so $M=M(x)$. It follows that the $vr$ component of equation (\ref{Richeq}) implies (see~\cite{KLR})
\be
Q''(x)+2\alpha^2=0
\ee
where $\alpha = L / \ell$, and thus using the boundary conditions $Q(\pm x_0)=\Gamma(\pm x_0) \gamma(\pm x_0)=0$ we get
\be
Q(x)=\alpha^2(x_0^2-x^2) \;.
\ee
The scaling symmetry (\ref{scaling1}) still allows some freedom in the definition of $x$. Noting that $Q \to s^{-2} Q$ and choosing $s= x_0^{-1} $ we get
\be
Q(x)=\alpha^2(1-x^2)
\ee
with  $-1\leq x \leq 1$. Therefore the $dx^2$ part of the near-horizon geometry metric is simply
\be
\frac{\ell^2\Gamma(x)dx^2}{(1-x^2)}
\ee
and
\be
\label{gamma}
\gamma =\frac{\alpha^2 (1-x^2)}{\Gamma(x) }   \; .
\ee

Now lets turn to the equation of motion for the matrix of scalar fields $M$. Since this only depends on the coordinate $x$ it becomes simply
\be
\frac{d}{dx} \left[ (1-x^2)M^{-1} \frac{d M}{dx} \right]=0  \; .
\ee
We can now exploit the results of~\cite{HI} where a useful form for the general solution to equations of this form was derived. First one can integrate to get
\be
M(x)=M(0) \left( \frac{1+x}{1-x} \right)^N
\ee
where $N$ is a constant matrix\footnote{We use the standard notation that $a^N\equiv \exp(N \log a)$.}. Now since $M$ is hermitian, positive and unimodular, essentially the same argument as in~\cite{HI} can be applied, which we now repeat. It is clear that $M(0)$ satisfies the same properties as $M(x)$,   $\textrm{Tr} N=0$ and $N^\dagger M(0)= M(0)N$. Also since $M$ is positive one can introduce a matrix $S$ such that $M(0)=S^\dagger S$ and $| \det S \, |= 1$. Thus $SNS^{-1}$ is a hermitian constant matrix, and by transforming $S\to VS$ for some unitary matrix $V$  we can always diagonalise $SNS^{-1}$ (note this does not change $M(0)$ and is thus a freedom in the definition of $S$ we may exploit). Thus we can write
\be
M_{IJ}(x)= \left[S^\dagger \left( \frac{1+x}{1-x} \right)^{SNS^{-1}}S\right]_{IJ}= \sum_{K=1}^n \left( \frac{1+x}{1-x} \right)^{\sigma_K} S^*_{KI} S_{KJ}  
\ee
where $\sigma_I$ are the (real) eigenvalues of $SNS^{-1}$. This expresses the matrix $M(x)$ as an explicit function of $x$ and depends on the parameters $S_{IJ}, \sigma_I$. Note that we have the constraints on the parameters
\be
\label{const}
| \det S \, |=1, \qquad \sum_{K=1}^n \sigma_K=0  \; .
\ee
Finally we may impose the $xx$ component of (\ref{Richeq}). Using~\cite{KLR} we get\footnote{This equation does not seem to appear in~\cite{HI}, although after a regularity analysis it turns out to be automatically satisfied for them. }
\be
\label{sigmasqsum}
2m=\textrm{Tr} N^2 = \sum_{K=1}^n \sigma_K^2 \; .
\ee
 We will show that a regularity analysis completely fixes these eigenvalues in our general setting. First note that for the 5d vacuum case one can show that the eigenvalues of the analogous $3\times 3$ matrix $N$ are $1,-1,0$~\cite{HI}. In 4d the analogous $2 \times 2$ matrix $N$ has eigenvalues $1,-1$. This suggests that the eigenvalues of our $n\times n$ matrix $N$ are $m$ pairs of $1,-1$ with the rest vanishing. Note that this solves both constraints on the eigenvalues. We now show the eigenvalues are also given by this in the non-vacuum case.

\subsection{Global analysis}
We will now perform a global analysis of the horizon metric. Recall the coordinate $-1\leq x \leq 1$ parameterises the orbit space $\hat{\mathcal{H}} = \mathcal{H}/ U(1)^{D-3}$ and we normalised the angles so that $\phi^i \sim \phi^i +2\pi$.  The analysis splits depending on whether $D=4$ or $D=5$.

In the $D=4$ case the horizon metric is simply
\be
\gamma_{ab} dx^a dx^b = \ell^2 \left[   \frac{ \Gamma(x) dx^2 }{1-x^2} + \frac{(1-x^2) d\phi^2}{\Gamma(x)} \right]
\ee
which is clearly invertible for $-1<x<1$. At the endpoints $x =\pm 1$ the Killing field $m=\partial / \partial \phi$ vanishes and absence of conical singularities at these points simply requires   $\Gamma_+ = \Gamma_-=1$. With these conditions our horizon metric is smooth and invertible with $\mathcal{H} \cong S^2$.

For $D=5$ the analysis is a little more involved  and we will follow the analysis of~\cite{HI}. At the boundary points $x=\pm 1$ certain integer linear combinations of the rotational Killing fields $m_i$ must vanish:
\be
\left.  v_{\pm}^i m_i  \right|_{x=\pm 1}=0
\ee
where $v^i_{\pm} \in \mathbb{Z}$ and we assume that gcd $v^i_{+}=$ gcd $v^i_-=1$. This is equivalent to
\be
\gamma_{ij}(x) v^i_{\pm} \to 0 \qquad x\to \pm 1  \; .
\ee
Thus the metric $\gamma_{ij}$ is degenerate at the endpoints -- however only one linear combination of the Killing fields vanishes at each point on the boundary and thus $\gamma_{ij}$ must be non-zero and rank-1, i.e.
\be
\gamma_{ij}(\pm 1) = w_i^{\pm} w_j^{\pm}   \label{wdef}
\ee
where $w^{\pm}_i$ are non-zero constant vectors. Notice $w^{\pm}_i v^i_{\pm}=0$.
In the interior $-1<x<1$ the 2-metric $\gamma_{ij}(x)$ is positive-definite and thus invertible. In fact using (\ref{Qdef}) we see that
\be
\label{gammaasy}
\frac{1}{\gamma} = \frac{\Gamma_{\pm}}{\alpha^2(1-x^2)} + O(1) \qquad \qquad x \to \pm 1
\ee
where $\Gamma_{\pm}= \Gamma(\pm 1)>0$ are positive constants. The horizon metric is
\be \label{3dmet}
\gamma_{ab}dx^a dx^b =L^2 \left( \frac{\Gamma(x) dx^2}{\alpha^2(1-x^2)} + \gamma_{ij}(x) d\phi^i d\phi^j \right)
\ee
which is invertible for $-1<x<1$ and generically has conical singularities at the endpoints $x=\pm 1$. Removal of these conical singularities is equivalent to:
\be
\label{conical}
 \gamma_{ij}v^i_{\pm} v^i_{\pm} = \alpha^{-2} \Gamma_{\pm} (1-x^2) + O[(1-x^2)^2] \qquad x \to \pm 1    \; ,
\ee
which can be thought of as determining $\alpha$.
With these conditions the horizon metric is everywhere smooth with a topology depending on the vectors $v_{\pm}$. Using the $SL(2,\mathbb{Z})$ freedom associated with the definition of the Killing fields $\partial / \partial \phi^i$ we may always choose $v_+ =(1,0)$. The topologies are then given in Table 1.
\begin{table}[!h]
\label{table}
\centering
\begin{tabular}{ | c | c | c | c | }
\hline
$\mathcal{H}$ & $S^1\times S^2$ &  $S^3$ & $L(p,q)$ \\
\hline
 $v_-$  & $(1,0)$ & $(0,1)$  & $(p,q)$ \\
 \hline
 \end{tabular}
 \caption{Possible horizon topologies in 5d}
\end{table}

We have just discussed the behaviour of the scalars $\gamma_{ij}$ as $x \to \pm 1$. The rest of the scalars are easier to examine at the endpoints. Since they only depend on $x$, regularity requires they must all tend to constants
\be
Y_i \to Y^\pm_i \qquad b^I_i \to b^{I \pm}_i \qquad \mu_I \to \mu_I^\pm
\ee
as $ x \to \pm 1$.

We now use these regularity conditions to deduce the eigenvalues of $N$ as promised. Since $M$ is built out of various potentials associated to the near-horizon geometry it must be the case that $\gamma^p M_{IJ}$ must be regular at the endpoints for some positive integer $p$ -- technically we will assume these matrix components are smooth functions.  We also assume $\Gamma(x)>0$ and is smooth. It follows that $(1-x^2)^pM_{IJ}(x)$ should also be smooth for all $-1\leq x\leq 1$ including the endpoints. From the solution for $M$ we have
\be
(1-x^2)^p M_{IJ}(x) = \sum_{K=1}^n (1+x)^{p+\sigma_K}(1-x)^{p-\sigma_K} S^*_{KI}S_{KJ}  \; .
\ee
Consider $S_{KI}$ for some fixed $K$. It must be the case that there is at least one $I$ such that $S_{KI} \neq 0$, otherwise $S_{KI}=0$ for all $I$ which implies $\det S=0$ a contradiction.  Since $(1-x^2)^p M_{IJ}$ must have a well defined limit as $x \to \pm 1$ it follows $-p\leq \sigma_K \leq p$. One might worry that there could be singular terms for different values of $K$ which cancel against each other; however, the only way this can occur is if there are two eigenvalues $\sigma_{K}=\sigma_{K'}$ and thus $(1-x^2)^p M_{II}(x) \sim (|S_{K I}|^2+|S_{K' I}|^2) (1+x)^{p+\sigma_{K}}(1-x)^{p-\sigma_{K}}$ which is necessarily singular (i.e. the coefficients of the singular terms are both positive and thus can never cancel) unless $-p\leq \sigma_K \leq p$. Furthermore, smoothness of $(1-x^2)^p M_{IJ}$ at the endpoints implies $\sigma_K$ is an integer. This argument is valid for each $K$ and thus we deduce that all the eigenvalues can only take the values $-p, \dots , -1,0,1 \dots, p$.

It turns out that in all known examples of theories in 4d and 5d the integer $p=1$. We will restrict to this case here.\footnote{It would be interesting to understand how much of a restriction this actually is.} Therefore each $\sigma_K$ can only be $-1,0,1$.  Since the sum of all eigenvalues vanishes we deduce that there must be an equal number of $+1$ and $-1$ eigenvalues with the rest vanishing. From (\ref{sigmasqsum}) we deduce that the multiplicity of both the $+1$ and $-1$ eigenvalues is $m$, with the rest vanishing.

To summarise we have shown that the general solution for $M$ can be written as:
\begin{empheq}[box=\fbox]{align}
\label{Msoln}
M_{IJ}(x)= \sum_{A=1}^m S^*_{AI} S_{AJ} \left( \frac{1+x}{1-x}\right)+ \sum_{B=m+1}^{2m} S^*_{BI} S_{BJ} \left( \frac{1-x}{1+x}\right)+\sum_{K=2m+1}^n S^*_{KI}S_{KJ} \; .
\end{empheq}
This expression completely fixes the functional form of the scalars/potentials and therefore it completely fixes the functional form of the near-horizon data. This essentially solves the classification problem for near-horizon geometries in a wide class of theories. We will illustrate this more explicitly for some sample theories.

Finally note that we may deduce the functional form for $\Gamma(x)$ from this general solution. It turns out that typically a linear combination of $M_{IJ}$ gives $\gamma^{-1}$, i.e. $\gamma^{-1} = L_{IJ} M_{IJ}(x)$ for some constants $L_{IJ}$. Combining this with (\ref{gamma}) determines $\Gamma(x)$ to be a quadratic function of $x$:
\be
\Gamma(x) =\frac{\Gamma_+\, (1+x)^2}{4} +\frac{\Gamma_- \,(1-x)^2}{4} + \Lambda\, (1-x^2)
\ee
where the constants $\Gamma_\pm, \Lambda$ can all be written in terms of $\alpha, S_{IJ}, L_{IJ}$.

\subsection{Physical charges}
In this section we give explicit formulas for the various physical charges and horizon area, for spacetimes containing extremal Killing horizons whose near-horizon geometries are of the form we have derived, purely in terms of the potentials. We note that in this section our formulas are valid for any near-horizon geometry (\ref{canonical}) in the general theory (\ref{gentheory}) with $U(1)^{D-3}$ rotational symmetry (i.e they do not require the sigma-model form of the equations of motion).

The area is the simplest to calculate (due to our choice of coordinates):
\be
A_H = \int_{\mathcal{H}} \epsilon_\gamma = \int_{\mathcal{H}}  L^{D-2} dx d\phi^1 \cdots d\phi^{D-3}= 2 (2\pi)^{D-3} L^{D-2}
\ee
which shows that $L$ is an invariant quantity of a solution.

We define the electric charge by the conserved charge at spatial infinity:
\begin{equation}
Q_I= \frac{1}{4\pi G} \int_{S_{\infty}}  (g_{IJ}(\phi) \star F^J - S_I) \; .
\end{equation}
Note that in four dimensions this expression is gauge invariant, as is the case in five dimensions assuming $F^I \to 0$ at spatial infinity. Using the Maxwell equation, one can then give an expression for this charge as an integral over $\mathcal{H}$:
\bea
Q_I &=& \frac{1}{4\pi G}\int_\mathcal{H}  \left( g_{IJ}(\phi) \star F^J -S_I  \right) \\
   &=&  \frac{1}{4\pi G} \int_{\mathcal{H}}  g_{IJ}(\phi) \Delta^J \, \epsilon_{D-2} - \begin{cases}
  \frac{1}{4}h_{IJ}(\phi)F^J & \text{if } D =4 \\
  - \frac{1}{8}C_{IJK}A^J \wedge F^K       & \text{if } D=5
 \end{cases}
\eea
where in the second line we have written the integrand explicitly in terms of the near-horizon data.  In four dimensions this may be evaluated explicitly using (\ref{NHmu}) and one finds
\begin{equation}
Q_I =\frac{L}{2G} (\mu_I^+ -\mu_I^-)
\end{equation}
where $\mu^\pm_I= {\mu_I}|_{x=\pm 1}$.
In five dimensions, one must take care to define the second term $\int_{\mathcal{H}} C_{IJK}A^J \wedge F^K$ properly, since in general we cannot assume the existence of a globally defined gauge field $A^I$ (in particular for $\mathcal{H} \cong S^1\times S^2$). For the topologies of interest it is sufficient to assume one can cover $\mathcal{H}$ with two charts such that $A^I$ can be made regular in each patch. Thus, split $\mathcal{H} = \mathcal{H}^+ \cup \mathcal{H}^- \cup E$ where $\partial \mathcal{H}^\pm = \pm E$ and $A^I_{\pm}$ is regular in $\mathcal{H}^\pm$ and $A^I_+-A^I_-=\beta^I$. Then the correct definition is~\cite{HOT}\footnote{This is to ensure that $\int_{\mathcal{H}} A \wedge F = \int_\Sigma F \wedge F$ where $\Sigma$ is any manifold such that $\partial \Sigma =\mathcal{H}$.}
\be
\int_\mathcal{H} C_{IJK} A^J \wedge F^K  \equiv \int_{\mathcal{H}^+}  C_{IJK} A^J_+ \wedge F^K + \int_{\mathcal{H}^-}  C_{IJK} A^J_- \wedge F^K + \int_E  C_{IJK} A^J_+ \wedge \beta^K  \; .
\ee
For our class of near-horizon geometries we can take $\mathcal{H}^\pm = \{ 0 \leq \pm x \leq 1 \}$ and $E= \{ x=0 \}$. We find that the charge formula then simplifies to
\be\label{Q5d}
Q_I=\frac{ \pi L^2}{G}\left[  \frac{1}{8} C_{IJK} (b_2^{J +} b_1^{K -} -b_1^{J +}b_2^{K -} )+ \int^1_{-1} dx  \left( L g_{IJ} \Delta^J- \frac{1}{8}C_{IJK}(b_1^J{b_2^K}'-b^J_2{b^K_1}') \right)   \right]
\ee
and thus using (\ref{NHmu}) we find
\be
\label{NHQ}
Q_I= \frac{\pi L^2}{G} \left[ \mu_I^+-\mu_I^- +\frac{1}{8} C_{IJK} (b_2^{J +} b_1^{K -} -b_1^{J +}b_2^{K -} ) \right]  \; .
\ee
One can check that this expression is in fact invariant under the gauge transformations (\ref{gauget1}) and (\ref{gauget2}).

Let us now consider magnetic charges. In four dimensions this is a genuine conserved charge given by
\begin{equation}
P^I = \frac{1}{4\pi G} \int_{S_\infty}  F^I  =\frac{1}{4\pi G} \int_\mathcal{H}  F^I
\end{equation}
where the second equality follows from using the Bianchi identity. This can be evaluated explicitly in terms of our potentials
\begin{equation}
P^I = \frac{L}{2G} (b^{I +} -b^{I -})  \; .
\end{equation}
In five dimensional asymptotically flat spacetimes there is no magnetic conserved charge. The closest analogue is the dipole charge which only exists for $\mathcal{H}=S^1\times S^2$ and is given by
\begin{equation}
\mathcal{D}^I = \frac{1}{2\pi} \int_{S^2} F^I
\end{equation}
where $S^2$ is the 2-sphere on $\mathcal{H}$. This evaluates to
\begin{equation}
\label{NHD}
\mathcal{D}^I =  L (b^{I +}_i  -b^{I -}_i)v^i
\end{equation}
where $v^i\partial_{\phi^i}$ is the Killing field with fixed points on the $S^2$, whose orbits are normalised to period $2\pi$.

One may also write down analogous formulas for the angular momenta, by using the Einstein equations to write the Komar integral as an integral over $\mathcal{H}$. This is straightforward in $D=4$. However, in $D=5$ for $S^1\times S^2$ topology, one again has to be careful to define the integral over $\mathcal{H}$ as the integrand again contains non-gauge invariant terms involving $A^I$. It would be interesting to do this in our general setup, as was done in~\cite{HOT}  for some specific examples. We will not pursue this here though.

\section{Example: $\mathcal{N}=1$, $D=5$  minimal supergravity}

\label{sec:5dsugra}
\subsection{General near-horizon geometry solution}
The bosonic field content of $D=5$ minimal ungauged supergravity consists of a metric tensor $g$ and a Maxwell 2-form $\mathcal{F}$. The field equations for this theory\footnote{We follow the conventions of \cite{Gauntlett} with the opposite signature.} are
\begin{eqnarray}\label{5EOM}
R_{AB} &=& 2 \mathcal{F}_{A}^{\phantom{A}C}F_{B C} - \frac{1}{3}g_{AB}\mathcal{F}^2 \\
 d\star_5 \mathcal{F} &+& \frac{2}{\sqrt{3}} \mathcal{F} \wedge \mathcal{F} = 0, \qquad d\mathcal{F} = 0. \label{5Max}
\end{eqnarray} These equations follow from a particular case of the general action (\ref{gentheory}) with one gauge field $I=1$, $g_{IJ}(\phi) =1$, $C_{IJK} = 16/\sqrt{3}$ as well as $V(\phi) =0$ and constant scalar fields.

We will restrict to solutions $(g, \mathcal{F})$ with a $U(1)^2$ isometry with spacelike orbits generated by Killing fields $m_i$ where $i=1,2$. As shown earlier this allows one to represent the Maxwell field in terms of scalar potentials.
It is convenient to rescale these potentials as
\begin{equation}\label{rescalemin}
\mu^I = \frac{\sqrt{3} \mu}{2} \qquad b_i^I = \frac{\sqrt{3} b_i}{2}  \; .
\end{equation}
We use the definitions of $\gamma_{ij}$ and the twist potentials $Y_i$ from the earlier general analysis.  Therefore in total we have 8 scalars $(\gamma_{ij} ,Y_i, b_i ,\mu)$. Note that the gauge transformation properties of these potentials may be deduced from (\ref{gauget1}), (\ref{gauget2}) and (\ref{gauget3}).

Remarkably, solutions to minimal ungauged supergravity with the $U(1)^2$ symmetries can be derived from a 3d sigma model defined on $\mathcal{M}_3$ of the form (\ref{Richeq})-(\ref{Meq}) with $m=2$. The target space of this sigma model can be identified with the coset space $G_{(2,2)}/SO(4)$. Therefore the matrix $M$ must be a representative of this coset and a convenient choice is as a $7\times7$ real symmetric matrix with unit determinant, which is positive definite (this follows from the fact we are reducing on spacelike Killing fields)~\cite{TYI}.  In the following, we will follow the presentation in \cite{TYI}, although we use our own notation above and make some minor changes.\footnote{To relate our potentials to  the analogous potentials used in~\cite{TYI}, note that their Maxwell field is twice ours, i.e. $F_{there}=2F_{here}$. Taking this into account we have the correspondence:
\bea
b_i=\psi_i, \qquad \mu_{here}=\mu_{there}, \qquad \gamma_{ij}= \lambda_{ij} \qquad Y_i=\omega_i
\eea}
Explicitly, the matrix $M$ can be written as
\be
\label{Mexplicit}
M= \left(
\begin{array}{ccc}  A & B & \sqrt{2} U \\
 B^T & C & \sqrt{2} V \\
 \sqrt{2} U^T & \sqrt{2} V^T & S \end{array} \right)
\ee
where $A,C$ are symmetric $3\times 3$ matrices, $B$ is a $3\times 3$ matrix, $U,V$ are $3\times 1$ matrices and $S$ is a scalar. These are given by
\bea
 \nonumber
S &=& 1 + 2(b_k b^k + \gamma^{-1}\mu^2) \\ \nonumber
U &=& \left( \begin{array}{c} (1+b_k b^k)b_i - \frac{\mu}{\sqrt{\gamma}}\epsilon_{i}^{\phantom{i}k}b_k + \frac{\mu}{\gamma}\tilde{Y}_i \\ \nonumber
-\frac{\mu}{\gamma} \end{array}\right) \nonumber \\
V &=& \left( \begin{array}{c} (\gamma^{ij} - \frac{\mu}{\sqrt{\gamma}}\epsilon^{ij})b_j \\ \gamma^{kl}b_k\tilde{Y}_l - \mu[1 + b_k b^k + \frac{\mu^2}{\gamma} - \frac{\epsilon^{kl}}{\sqrt{\gamma}}b_k \tilde{Y}_l] \end{array} \right) \nonumber \\
A &=& \left( \begin{array}{cc}
(1+\gamma^{-1} \mu^2)\gamma_{ij} + \gamma^{-1} \tilde{Y}_i \tilde{Y}_j +(2+ b_kb^k)b_ib_j+\frac{\mu}{\sqrt{\gamma}} (b_i b^k \epsilon_{kj} -\epsilon_{ik}b^k b_j ) & - \gamma^{-1} \tilde{Y}_i \\
-\gamma^{-1}\tilde{Y}_j & \gamma^{-1} \end{array} \right) \nonumber   \\
B & = & \left(\begin{array}{cc} b_i b^j  - \frac{\mu}{\sqrt{\gamma}}\epsilon_{i}^{\phantom{j}j} + \frac{1}{\sqrt{\gamma}}\tilde{Y}_i b_k \epsilon^{kj} & \beta_i \\
-\frac{b_k \epsilon^{kj}}{\sqrt{\gamma}} & \frac{\mu^2}{\gamma} - \frac{\epsilon^{lm}b_l \tilde{Y}_m}{\sqrt{\gamma}} \end{array}\right)
\\ \nonumber
C &=& \left( \begin{array}{cc}   (1+b_kb^k)\gamma^{ij} -b^ib^j & \tilde{Y}^i +(\frac{\mu^2}{\sqrt{\gamma}} - b_l \tilde{Y}_m \epsilon^{lm})  \epsilon^{ik}b_k -\mu b^i  \\  \tilde{Y}^j +(\frac{\mu^2}{\sqrt{\gamma}} - b_l \tilde{Y}_m \epsilon^{lm})  \epsilon^{jk}b_k -\mu b^j   & c \end{array} \right)
\eea
where we raise and lower all indices with the metric $\gamma_{ij}$ and have defined
\bea
\tilde{Y}_i &=&Y_i-\mu b_i  \\
\beta_i &=& -\left(1 - \frac{\mu^2}{\gamma}\right)\sqrt{\gamma}\epsilon_{i}^{\phantom{i}k}b_k  - (2 + b_k b^k)\mu b_i + \gamma^{kl}b_k \tilde{Y}_l b_i  + \left(-\frac{\mu^2}{\gamma} + \frac{\epsilon^{kl}b_k \tilde{Y}_l}{\sqrt{\gamma}}\right)\tilde{Y}_i  - \frac{\mu}{\sqrt{\gamma}}\epsilon_{ik}\tilde{Y}^k \nonumber \\
 c &=& \tilde{Y}^k\tilde{Y}_k - 2\mu b^k\tilde{Y}_k +\gamma[1+ b_kb^k+(2+b_kb^k)\gamma^{-1}\mu^2 +\gamma^{-2}(\mu^2- b_l \tilde{Y}_m \sqrt{\gamma}\epsilon^{lm} )^2 ]   \nonumber  \; .
\eea
Now let us introduce some notation regarding the matrix indices of $M$. We will refer to a general component by $M_{IJ}$ where $I,J =1, 2, \dots, 7$. An index $I=(i, 3, j+3, 6, 7)$ where as always $i,j=1,2$, so for example
\bea
&&M_{ij}=A_{ij} \qquad M_{i3}=A_{i3} \qquad M_{33}=A_{33} \\
&&M_{i \, j+3}= B_{i}^{\phantom{j}j} \qquad M_{i6}=\beta_i  \\
&&M_{i+3 \, j+3}= C^{ij}  \qquad M_{66}=c \\
&& M_{i7}= \sqrt{2} U_i \qquad M_{i+3 , 7} =\sqrt{2} V^i \qquad M_{77}=S   \; .
\eea
It is useful to show how the vacuum case is embedded in this formalism. If one sets $b_i=\mu=0$ we see that $A=\chi$, $S=1$, $U=V=0$, $B=0$ and $C=\chi^{-1}$, where $\chi$ is the vacuum $SL(3,R)$ matrix (\ref{chi}),
and thus
\be
M= \left(
\begin{array}{ccc}  \chi & 0 & 0 \\
 0 & \chi^{-1} & 0 \\
 0 & 0 & 1 \end{array} \right)
\ee
Note that using $A^{-1}dA=-A dA^{-1}$ it is easy to see that Tr$(M^{-1}dM)^2= $2Tr$(A^{-1} dA)^2$ and thus the above reduces correctly to the vacuum equations $R^{(3)}_{\mu\nu}= \frac{1}{4}$Tr$(\chi^{-1} \partial_\mu \chi)(\chi^{-1} \partial_\nu \chi)$.

Finally, we can now identify the 3d data associated with near-horizon solutions of minimal five-dimensional supergravity with two commuting spacelike Killing fields  as discussed in Section \ref{general}. It is convenient to rescale as
\begin{equation}
e^I = \frac{\sqrt{3} e}{2}, \qquad \Delta^I = \frac{\sqrt{3} \Delta}{2},
\end{equation} in which case
\begin{eqnarray}
\mu' &=& L \Delta - (b_1 b_2' - b_2 b_1') \;, \\
Y_i'&=& \frac{\gamma_{ij}k^j}{\Gamma} + b_i \left[3 \mu' + (b_1 b_2' - b_2 b_1')\right] \; .
\end{eqnarray} Note that given the scalars $(\gamma_{ij}, Y_i, b_i, \mu)$ we may invert the above relations to determine the near-horizon data $(k^i, \Delta)$:
\bea
\label{NHdata1}
&&k^i= \Gamma \gamma^{ij} \left[ Y_j' -b_j\left( 3 \mu' +(b_1b_2'-b_2b_1')  \right) \right]  \\
&&L \Delta =   \mu' +(b_1b_2'-b_2b_1') \; . \label{NHdata2}
\eea

We now proceed to finding the general functional form of all near-horizon geometry solutions of minimal five-dimensional supergravity by equating our general solution~(\ref{Msoln}) (with $m=2$ and $n=7$) and the explicit representation of $M$ of~(\ref{Mexplicit}). Since $M$ is symmetric and unimodular it has 27 independent components and therefore this is the maximum number of equations for the 8 potentials. Therefore this algebraic system is highly overdetermined, which leads to additional constraints on the $S_{IJ}$.

The 33 component of this equation determines $\gamma(x)$ and hence from (\ref{gamma}) yields
\begin{equation}
\Gamma(x) = \frac{\Gamma_+ (1+x)^2}{4} + \frac{\Gamma_- (1-x)^2}{4} + \alpha^2(1-x^2) \sum_{K=5}^7 S_{K3}^2
\end{equation} where
\begin{equation}
S_{13}^2 + S_{23}^2 = \frac{\Gamma_+}{4\alpha^2}, \qquad S_{33}^2 + S_{43}^2 = \frac{\Gamma_-}{4\alpha^2}
\end{equation} so that $\Gamma(x)>0$ as required.
The $i3$ and $37$ components of (\ref{Mexplicit}) then give
\begin{equation}
\tilde{Y}_i = \frac{ \Gamma_+ \tilde{Y}^+ _i(1+x)^2 + \Gamma_- \tilde{Y}_i^-(1-x)^2 - 4\alpha^2(1-x^2)\sum_{K=5}^7S_{Ki}S_{K3}}{4\Gamma(x)}
\end{equation}
and
\be
\mu =  \frac{ \Gamma_+ \mu_+ (1+x)^2 +\Gamma_- \mu_- (1-x)^2 - 2\sqrt{2}\alpha^2(1-x^2) \sum_{K=5}^7 S_{K3}S_{K7} }{4\Gamma(x)}
\ee
respectively, where
\bea
&&\tilde{Y}_i^+ = -\frac{S_{1i}S_{13}+S_{2i}S_{23}}{S_{13}^2+S_{23}^2} \qquad \tilde{Y}_i^- = -\frac{S_{3i}S_{33}+S_{4i}S_{43}}{S_{33}^2+S_{43}^2}  \label{Ypm}  \\
&& \mu_+ = - \frac{S_{13}S_{17}+S_{23}S_{27}}{\sqrt{2}(S_{13}^2+S_{23}^2)} \qquad \mu_-= - \frac{S_{33}S_{37}+S_{43}S_{47}}{\sqrt{2}(S_{33}^2+S_{43}^2)}  \label{mupm}
\eea Next, the $3 \; 3+i$ component of (\ref{Mexplicit}) can be solved for the potential $b_i$ giving:
\be \label{bdown}
b_i = \frac{ \Gamma_+ b_i^+ (1+x)^2 +\Gamma_- b^-_i (1-x)^2  -4\eta_{ij} \alpha^2(1-x^2) \sum_{K=5}^7 S_{K3}S_{K}^{\phantom{K} j}}{4\Gamma(x)}
\ee
where $\eta_{ij}=\epsilon_{ij}/\sqrt{\gamma}$ is a tensor density ($\eta_{12} = +1$), we have used the notation $S_{1 j+3}= S_{1}^{\phantom{1} j}$ etc in order to maintain manifest covariance, and
\be
b_i^+ =-\frac{\eta_{ij}(S_{13}S_1^{\phantom{1}j}+S_{23}S_2^{\phantom{1}j})}{S_{13}^2+S_{23}^2} \qquad b_i^- =-\frac{\eta_{ij}(S_{33}S_3^{\phantom{1}j}+S_{43}S_4^{\phantom{1}j})}{S_{33}^2+S_{43}^2}    \; .
\ee It is clear that these expressions for ($\mu, b_i, Y_i$) are regular at the endpoints, as required by our general analysis above.  The final potentials to determine are the horizon metric components $\gamma_{ij}$. To do so, it proves convenient to first find expressions for $b^i = \gamma^{ij} b_j$ and thus $b^i b_i$.   From the $i+3 \,  7$ component of (\ref{Mexplicit}) we find
\begin{eqnarray} \label{bup}
b^i = \frac{M_{i+3 \, 7}}{\sqrt{2}} + \mu M_{3 \, 3+i}  \; .
\end{eqnarray}
The final scalars to be determined are the metric components $\gamma_{ij}$. These can now be reconstructed from the $ij$ component of (\ref{Mexplicit}) which gives
\be
\label{gammaij}
\gamma_{ij} = \frac{\gamma M_{ij} - \tilde{Y}_i\tilde{Y}_j - (2+ b_k b^k)\gamma b_i b_j + 2\mu \sqrt{\gamma}b_{(i}\epsilon_{j)k}b^k}{\gamma+\mu^2} \; .
\ee
We have explicit expressions for all quantities on the RHS and hence for $\gamma_{ij}$.
The formulas discussed so far thus give us the explicit $x$ dependence of the $8$ potentials and using (\ref{NHdata1}) and (\ref{NHdata2}) that of the near-horizon data. It is worth observing that $\gamma_{ij}$ are always rational functions of the geometrically defined coordinate $x$, in general a quotient of two octic polynomials.

Clearly, we have not yet used all components of (\ref{Mexplicit}), which will impose additional non-trivial constraints on the $S_{IJ}$.  For example, the 77 component requires
\be
\label{bsq}
b_kb^k = \frac{M_{77}}{2} - \frac{1}{2} - \frac{\mu^2}{\gamma}  \; .
\ee Evaluating this expression at the endpoints and comparing to the same quantity computed by contracting (\ref{bup}) with (\ref{bdown}) yields the constraints
\be
\label{constraint1}
\sqrt{2} \eta_{ij}S_{1}^{\phantom{1}i} S_{2}^{\phantom{5}j} =S_{13}S_{27}-S_{17}S_{23}  \qquad \qquad
\sqrt{2} \eta_{ij}S_{3}^{\phantom{1}i} S_{4}^{\phantom{4}j} =S_{33}S_{47}-S_{37}S_{43}  \; .
\ee
We have derived the full set of such constraints evaluated at the endpoints,  which take the form of higher-order polynomial equations relating the $S_{IJ}$.  We do not display them here as they are cumbersome expressions.  Nonetheless, we may still deduce general properties of our solutions. For example in the Appendix, these constraints can be used to show $\gamma_{ij}$ is rank-1 at the endpoints, as required by regularity.

\subsection{Examples of near-horizon geometries with $S^1 \times S^2$ horizon topology}
\label{example}

In this subsection we present some of the known examples of near-horizon geometries with $S^1\times S^2$ horizon topology, together with their potentials which all take the general functional form we have derived above (as they should).\\

\noindent {\it Five-parameter extremal black string} \\
\noindent Now we present a five-parameter solution, describing the non-static\footnote{For examples of static near-horizon geometries of this theory, see \cite{KLstatic}. } near-horizon geometry of a five-parameter extremal black string, whose full solution was first constructed in \cite{CBSV}.  This black string solution carries independent linear momentum $P$ and angular momentum $J$ along the $S^1$ and $S^2$ of the string respectively, as well as electric charge $Q_e$ and magnetic charge $Q_m$. The 8 scalars $\gamma_{ij}, \mu, b_i, \tilde{Y}_i$ that fully determine the solution are conveniently parameterized by five constants : $(a, \beta, \delta, \gamma, R)$ (the final parameter corresponds to the radius of the $S^1$ at spatial infinity). The explicit solution, which is fairly cumbersome, is given in full in the Appendix. For clarify, here we exhibit only the \emph{functional} form of the near-horizon solution. The various constants and functions can be easily read off from (\ref{expsola})-(\ref{expsolb}). The horizon metric is given by (\ref{3dmet}) with
\begin{equation}\label{schemet}
 \gamma_{ij}(x) d\phi^i d\phi^j = \frac{\Gamma(x)}{P(x)}\left((1-x^2)\left[d\phi^2- q_0 d\phi^1 \right]^2  + \frac{1}{\Gamma(x)^3}\left[a(x) d\phi^1 +(1-x^2) b(x) d\phi^2\right]^2 \right)
\end{equation} where, as dictated by our general analysis, $\Gamma(x)$ is a quadratic function. The remaining metric functions are $P(x)$, a quartic,  and $a(x)$ and $b(x)$ which are respectively a quartic and a quadratic, and $q_0$ is a constant. Analysis of the local metric~(\ref{3dmet}, \ref{schemet}) shows that it extends smoothly to a cohomogeneity-one metric on $S^1 \times S^2$. The Killing vector field $\partial / \partial \phi_2$ vanishes on the poles of the $S^2$.  The remaining potentials take the following functional form
\begin{eqnarray}\label{schmet2}
&& b_i(x) = \frac{b_i^0 + b_i^1 x}{\Gamma(x)}, \qquad \mu(x) = \frac{\mu_0 + \mu_1 x}{\Gamma(x)} \\
&& \tilde{Y}_i(x) = Y_i(x) - \mu(x)b_i(x) = \frac{\tilde{Y}_i^0 + \tilde{Y}_i^1 x}{\Gamma(x)}  \; .
\end{eqnarray} We emphasize that the various constants defined above may be obtained straightforwardly from the explicit expressions in the Appendix. The form of the scalars is still a remarkable simplification given the complexity of the full black string solution \cite{CBSV}.  Finally, note that the locally AdS$_3 \times S^2$ near-horizon geometry (see below) arises as a limit of the above near-horizon geometry. Details are given in the Appendix.

A considerable simplification of this five-parameter solution occurs if one considers an extremal black string with vanishing electric charge. We will refer to this solution as the `magnetic black string'  \cite{CBJV} and its near-horizon limit can be found by setting $\delta=0$ in the Appendix (see also \cite{CWV}). The resulting four-parameter solution has scalars:
\begin{eqnarray}
\gamma_{ij}(x)d\phi^i d\phi^j &=& \frac{a^4 (1-x^2)}{L^2 \ell^2 \Gamma(x)} \left(2(c_\beta^4 + s_\beta^4)d\phi^2 - \frac{s_\gamma R}{a} d\phi^1 \right)^2  \\ &+& \frac{a^2}{L^2} \left[  a(x) d\phi^1 + \frac{2 a^2 c_\beta s_\beta(c_\beta^2 + s_\beta^2)(1-x^2)}{\ell^2\Gamma(x)}d\phi^2 \right]^2 \nonumber
 \end{eqnarray} where
 \begin{equation}
\Gamma(x) = \frac{a^2}{\ell^2}\left(1+x^2+4c_\beta^2s_\beta^2\right), \qquad  a(x) = \frac{R}{a}\left( c_\gamma - \frac{2 a^2 s_\gamma c_\beta s_\beta (c_\beta^2 + s_\beta^2)}{\ell^2 \Gamma(x)} \right) \; .
\end{equation} The $AdS_2$ and horizon length scales are
\begin{equation}
\ell^2 = 2 a^2\left[\left(c_\beta^4 + s_\beta^4\right)c_\gamma - c_\beta s_\beta (c_\beta^2+ s_\beta^2) s_\gamma\right], \qquad L^3 = R\ell^2  \; .
\end{equation} The remaining near-horizon scalars are
\begin{eqnarray}
&& b_1(x) = -\frac{2 R a^2 s_\gamma s_\beta c_\beta x}{L \ell^2 \Gamma(x)} ,  \qquad b_2(x) = \frac{4 a^3 s_\beta c_\beta (c_\beta^4 + s_\beta^4)x}{L \ell^2 \Gamma(x)}  \nonumber \\
&& \mu(x) = \frac{2 a L c_\beta s_\beta (c_\beta^2 + s_\beta^2)}{\ell^2 \Gamma(x)}
\end{eqnarray} The shifted twist potentials are
\begin{equation}
\tilde{Y}_1 = -\frac{2 R a  c_\beta s_\beta x }{\ell^2 \Gamma(x)}, \qquad \tilde{Y}_2 =  -\frac{4a^2(c_\beta^2 + s_\beta^2)(1+s_\beta^2 c_\beta^2) x}{\ell^2 \Gamma(x)}  \; .
\end{equation}  Note that from these potentials one can immediately compute the electric charge from (\ref{NHQ}) and find $Q=0$ as claimed, and also from (\ref{NHD}) generically the dipole charge $\mathcal{D} \neq 0$. The charges are explicitly given in the Appendix. \\

\noindent {\it Supersymmetric Black Ring}

\noindent For comparison, we give the corresponding scalars for the near-horizon geometries of other known branches of extremal black rings. The near-horizon limit of the supersymmetric black ring \cite{EEMR}  is simply (locally) $AdS_3 \times S^2$ with
\begin{eqnarray}
&& \gamma_{ij} d\phi^i d\phi^j = \frac{R^2}{L^2} (d\phi^1)^2 + (1-x^2)(d\phi^2)^2 \\
&& \Gamma = 1, \qquad b_2 = x, \qquad Y_1 = \frac{R x}{L}
\end{eqnarray} with $\ell=L$ and the remaining scalars vanish. The dipole charge in this case is simply $\mathcal{D}=\sqrt{3} L$. \\

\noindent {\it Extremal Dipole Ring}

\noindent The extremal dipole ring\cite{emparan} has a three-parameter near-horizon geometry\cite{KLR} parameterised by $(q,\lambda,R_1,R_2)$ with one constraint between them\footnote{Regularity of the full dipole ring\cite{emparan} fixes the parameter $R_1$, but this is not required for smoothness of the near-horizon geometry, so we leave it free.} with scalars:
\begin{eqnarray}
\label{dipoleringNH}
\gamma_{ij}(x)d\phi^i d\phi^j &=& \frac{R_1^2\lambda(1+\lambda) H(x)}{q L^2 (1-\lambda)F(x)}(d\phi^1)^{2} + \frac{R_2^2q^2 \omega_0^2 (1-x^2)}{L^2 H(x)^2} (d\phi^2)^2     \\
\Gamma(x) &=& \sqrt{\frac{q(1-\lambda)}{\lambda(1+\lambda)}} \; F(x)H(x)
\end{eqnarray} where $F(x)=1+\lambda x$, $H(x)=1-q x$ and $0 < \lambda, q < 1$.  The local metric extends smoothly to a cohomogeneity-1 metric on $S^1 \times S^2$ provided conical singularities are removed, which requires
\be
\label{dipoleperiod}
\omega_0 = \sqrt{F(1)H(1)^3} =  \sqrt{F(-1)H(-1)^3}
\ee which imposes a relation between $(\lambda,q)$. The AdS$_2$ and horizon length scales are
\begin{equation}
\ell^2= R_2^2 \sqrt{\frac{\lambda(1+\lambda) q^{3}}{1-\lambda}}, \qquad L^3= \omega_0 R_1 \ell^2 \; .
\end{equation} The remaining non-vanishing near-horizon scalars are
\begin{equation}
b_2(x) = \sqrt{\frac{1-q}{1+q} } \; \frac{\omega_0 q R_2
(1+x)}{L H(x)} \; , \qquad Y_1(x) = \frac{R_1(1+\lambda)}{\sqrt{(\lambda q^3)} R_2} \left(1-\frac{1}{F(x)} \right)
\end{equation} and note that in particular $\mu = 0$. Using (\ref{NHQ}) and (\ref{NHD}), it is easily seen that $Q=0$ and $\mathcal{D} \neq 0$ as expected.

It is worth noting that the locally AdS$_3\times S^2$ near horizon geometry can be recovered as a limit of these near-horizon geometries. One sends $q,\lambda \to 0$ holding $q/ \lambda$, $R_2 q$ and $R_1$ fixed (so $R_2 \to \infty$). In fact this corresponds to the infinite radius limit of the parent dipole black ring, taken purely at the level of the near-horizon geometry.

The dipole ring \cite{emparan} has a charged generalisation~\cite{EEF} which in turn admits an extremal limit. The corresponding near-horizon geometry has a similar functional form to (\ref{dipoleringNH}) with $\mu(x) \neq 0$.

\section{Summary $\&$ the space of extremal black rings}
\label{discussion}
Constructing near-horizon geometry solutions in theories of gravity coupled to Maxwell and scalar fields (such as supergravity) is a difficult task. We have shown that for solutions which admit a $U(1)^{D-3}$ rotational symmetry, and the subset of such theories which are equivalent to a 3d theory of gravity coupled to a non-linear sigma model, it is possible to completely integrate the Einstein equations to find the general solution.  However, although the method provides an elegant means to determine the functional form of the solutions, the difficulty in solving the resulting set of algebraic constraints is a serious practical obstacle.  This prevents us at present from achieving an {\it explicit} classification of near-horizon geometries in theories of interest such as five dimensional minimal supergravity.

Our work was partially motivated by the fact that the most general black ring solution to five dimensional minimal supergravity remains to be found. Finding this could help clarify their microscopic description in string theory \cite{Larsen,emparan2}. We recall that currently, there are four separate known families of black ring solutions to minimal supergravity.\footnote{We will only consider asymptotically flat solutions with a single regular horizon. We also restrict our discussion to the minimal theory, although there are analogous solutions in the more general $U(1)^3$ supergravity. } There is the three parameter supersymmetric black ring $(J_1,J_2,Q)$ ($M=\sqrt{3}Q/2$ with $J_1>J_2$)~\cite{EEMR},  a three parameter singly spinning dipole black ring $(M,J_1, \mathcal{D})$~\cite{emparan}, a three parameter electrically charged black ring (this can never be supersymmetric)~\cite{EEF}, and the three parameter vacuum black ring $(M,J_1,J_2)$~\cite{PS}. Naturally, it has been conjectured that all these solutions are special cases of a five parameter family of black ring solutions which carry all five charges $(M,J_1,J_2, Q,\mathcal{D})$ independently~\cite{EEF}. We note that all the known solutions have $U(1)^2$ rotational symmetry and thus it is reasonable to expect such a general solution to as well.

Now, restricting attention to the extremal case, we deduce that extremal nonsupersymmetric black rings should carry \emph{four} charges $(J_1,J_2, Q, \mathcal{D})$ independently. However, it is entirely possible that there are multiple extremal limits which lead to distinct families of extremal black rings (and some of these could even have less than four parameters).\footnote{This would be analogous to the vacuum KK black hole which has two distinct extremal limits termed the fast and slow rotating cases.}

Such extremal black rings will have associated near-horizon geometries (with $U(1)^2$ rotational symmetry). These will either be already known, or correspond to new near-horizon geometry solutions to minimal supergravity. Therefore a natural problem is to actually classify all near-horizon solutions in five-dimensional minimal supergravity, admitting $U(1)^2$ rotational symmetry, with spatial horizon topology $S^1\times S^2$. This would therefore necessarily include the near-horizon geometries of these yet to be found extremal black rings.  However, one is then faced with identifying which near-horizon geometries actually correspond to asymptotically flat black rings, as opposed to black strings or black holes with  other types of asymptotic behaviour (such as background Maxwell fields~\cite{KLstatic}), which in general is a difficult inverse problem.  Although we did not quite achieve an explicit classification,  as discussed above we have in fact determined the functional form of any near-horizon geometry in this class.  This knowledge should be helpful in finding new explicit examples of near-horizon geometry solutions.

In fact, the largest explicitly known family of near-horizon geometries with spatial horizon topology $S^1\times S^2$ in minimal supergravity, is a five-parameter solution corresponding to the near-horizon limit of an extremal {\it boosted} black string\footnote{In fact it was shown in \cite{CBSV} there are two other possible extremal limits (see also \cite{nil}) but the associated near-horizon geometries are simply AdS$_3 \times S^2$.} found in \cite{CBSV}. In this paper,  we have constructed this rather complicated five parameter near-horizon geometry explicitly, see Appendix \ref{kerrstring}.  Now,  as discussed in \cite{CBSV}, the subset of {\it tensionless} black string solutions (which correspond to fixing the boost to some value) is expected to describe the infinite radius limit of a yet-to-be found black ring with $(M,J_1,J_2, Q, \mathcal{D})$ (whether it is extremal or not).   The tensionless condition for these extremal, charged black strings is given by (\ref{tensionzero}). Interestingly, it turns out\cite{CBSV} that there are two distinct values of the boost which solve this condition, thus leading to two four parameter families of near-horizon geometries. These correspond to the `aligned' and `anti-aligned' cases in the nomenclature of~\cite{CBSV}, referring to whether the linear momentum along the string is oriented with the magnetic charge or not.

On the other hand, as observed in \cite{KLR} and elaborated upon in \cite{FKLR}, the near-horizon geometry of the extremal vacuum doubly-spinning black ring is identical to that of the extremal tensionless Kerr string. Furthermore, the near-horizon geometry of the supersymmetric black ring, which is simply a quotient of AdS$_3 \times S^2$, is the same as that of a tensionless supersymmetric black string solution. In both of the above cases, the associated string is in fact \emph{also} the infinite radius limit of the corresponding black ring. Hence, for these two examples, the near-horizon geometry does not capture any finite radius effects. One might expect this to be true more generally. However, the singly spinning extremal dipole ring~\cite{emparan} provides a counterexample to this proposal. Its infinite radius limit is also in fact another tensionless, supersymmetric black string with associated near-horizon geometry locally AdS$_3 \times S^2$. This is clearly \emph{not} the inhomogeneous near-horizon geometry of the dipole ring (\ref{dipoleringNH}). We conclude that in this example, finite-radius effects are present in the near-horizon geometry (in section \ref{example} we show how to take the infinite radius limit purely at the level of the near-horizon geometry).

The remarkable fact that the near-horizon geometries of the extremal vacuum and supersymmetric black ring are the same of those of their infinite radius limits, can be understood as a consequence of the fact that in both of these classes there is a \emph{unique} near-horizon geometry with $S^1 \times S^2$ spatial horizon sections~\cite{Reall, KLvac}. Hence, finite radius effects which would distinguish the near-horizon geometries of black rings and black strings must be absent in these cases. However, in the absence of near-horizon geometry uniqueness we see there is no mechanism to prevent one from losing finite radius effects in the near-horizon geometry when one takes the infinite radius limit.  Within minimal supergravity we have seen that near-horizon geometries are parameterised by a finite number of constants (the constants $S_{IJ}$ subject to some number of constraints), but it is unclear what the exact number of parameters is for $S^1\times S^2$ horizon topology and in particular whether it is greater than four.

Let us now consider the near-horizon geometries of the four parameter tensionless extremal black strings discussed above.  The obvious question is whether these are isometric to the near-horizon geometries of the yet to be found four parameter extremal black rings with electric and dipole charge.  Indeed, setting $\delta=\beta=0$, so $Q=\mathcal{D}=0$, gives the near-horizon geometry of the extremal vacuum black ring (i.e. the near-horizon geometry of the extremal Kerr-string with its tensionless condition $\sinh^2\gamma=1$). Furthermore, the near-horizon geometry of the supersymmetric black ring, which is a quotient of AdS$_3\times S^2$, arises as a limit in the parameterisation we use (see Appendix \ref{kerrstring}). However, the other examples, namely the near-horizon geometries of the extremal dipole ring and its charged version, do not appear to be contained in our 4 parameter families, although their infinite radius limits (AdS$_3\times S^2$) are.  We are therefore led to two possibilities. One possibility is that these four parameter families of near-horizon geometries are the infinite radius limits of the actual black ring near-horizon geometries. Another possibility is that there is a branch of extremal black rings whose near-horizon limits coincide with those of their the infinite radius limits, and that the extremal dipole ring and its charged version belong to a distinct branch.

It order to get some insight into this, let us examine the $Q=0$ cases of our four parameter families of near-horizon geometries, so that they become three parameter families, which should correspond to (at least) the infinite radius limit of extremal black rings with independent charges $(J_1,J_2,\mathcal{D})$, i.e. doubly spinning dipole back rings. Note the only way to set $Q=0$, while keeping the string tensionless, is to set $\delta=0$ (see Appendix).\footnote{This can also be found as the near-horizon limit of the magnetically charged tensionless Kerr string which arises from setting $Q=0$ in the full solution (this string was first given in~\cite{CBJV} in a slightly different form).} It can be seen from the parameterisation given in \cite{CBJV} that the two solutions of the tensionless condition correspond to having equal and opposite momentum along the string (given all other parameters are fixed). In fact the extremal magnetic boosted Kerr string has a
simple expression for the area of the horizon as a function of
charges (for any boost, with $G=1$):
\be A_H = 8 \pi \sqrt{ J_{\phi'}^2 - \frac{2\pi}{3\sqrt{3}}
P_{\psi'}\mathcal{D}^3} \ee
where $P_{\psi'}$ is the momentum along the
string converted to the coordinate $\psi'=z/R$ and $J_{\phi'}$ is the angular momentum with respect to the $S^2$ Killing field. Recall that in the vacuum case
$\mathcal{D}=0$ the identification between the string angles $(\psi',\phi')$
and the black ring angles $(\psi,\phi)$ is $\phi'=\psi+\phi$ and
$\psi'=\psi$, where $\psi$ is the $S^1$ angle and $\phi$ the $S^2$ angle~\cite{KLR} (see also~\cite{FKLR}). By continuity this
must be true for $\mathcal{D} \neq 0$ and thus we must have
$J_{\phi'}=J_{\phi}$ and $P_{\psi'}=J_{\psi}-J_{\phi}$.   Now, although our near-horizon geometries do not necessarily include finite radius effects, in fact the area formula is typically insensitive to this \cite{emparan2}. We deduce that the branch of black ring solutions corresponding to our near-horizon geometries would always have a lower bound on $J_\phi$ given by $J_{\phi}^2>\frac{2\pi}{3\sqrt{3}}(J_{\psi}-J_{\phi}) \mathcal{D}^3$.  This branch would contain the known extremal vacuum black ring.

Given this it is tempting to speculate that the area formula for all extremal doubly spinning dipole black rings is given by
\be
A_H=8 \pi \sqrt{
\left| J_{\phi}^2 - \frac{2\pi}{3\sqrt{3}}(J_{\psi}-J_{\phi})\mathcal{D}^3 \right|}   \; .
\label{fastring}
\ee
This includes the above branch $J_{\phi}^2>\frac{2\pi}{3\sqrt{3}}(J_{\psi}-J_{\phi}) \mathcal{D}^3$, and also for $J_\phi=0$ reduces to the correct formula for the known singly spinning extremal dipole black ring $A_H = 8\pi \sqrt{
\frac{2\pi}{3\sqrt{3}} \left|J_{\psi} \mathcal{D}^3 \right|}$~\cite{emparan}.  Furthermore, this formula then suggests another branch of extremal black rings which have an upper bound on $J_\phi^2 < \frac{2\pi}{3\sqrt{3}}(J_{\psi}-J_{\phi}) \mathcal{D}^3$ (assuming that the full configuration space is filled).\footnote{This would be analogous to the fast and slow extremal limits of the 5d vacuum KK black hole for which the area formula $A= 8\pi \sqrt{| PQ-J|}$ in both cases.} This would contain the singly spinning dipole ring and generalisations thereof. It would be very interesting to test this idea by finding an explicit near-horizon geometry corresponding to this, which should be made easier by the general classification results presented in this paper.

In the presence of electric charge this picture becomes more complicated.  In fact, as shown in~\cite{CBSV}, the two solutions to the tensionless condition actually represent physically different solutions, i.e. for fixed $J_{\phi'}, \mathcal{D}, P_{\psi'}$  the electric charges are different. Therefore even in this general case there seem to be multiple branches of extremal black rings.
The status of the three parameter supersymmetric black ring is also unclear, since as mentioned above its near-horizon geometry (i.e. locally AdS$_3\times S^2$) arises as a limit of both of the 4 parameter near-horizon geometries we have presented here, and also as a limit of the near-horizon geometry of the known singly spinning dipole ring.  It thus appears that due to its high degree of symmetry, AdS$_3\times S^2$ always arises as a limit of such near-horizon geometries (provided they have a non-zero dipole charge). However, this does not imply that the supersymmetric black ring would be contained in all of the families of extremal black rings, but rather that a near-horizon analysis is insufficient to locate this solution.

To summarise, based on explicit examples of four parameter near-horizon geometries, we have argued that the space of extremal black rings in minimal supergravity is most likely not connected. Our arguments, built from studying near-horizon geometries with $S^1 \times S^2$ horizon topology, are complementary to the analysis of \cite{CBSV} which is based on considering the infinite radius limits of black rings and the blackfolds approach \cite{CEvP}.  One branch should contain the singly spinning dipole ring and the known charged black ring solutions (including perhaps the supersymmetric black ring), whereas another branch would contain the vacuum black ring (and perhaps the supersymmetric black ring). Furthermore, we have (at least) the infinite radius limit of the near-horizon geometry of the latter branch of extremal black rings. It would be most interesting to construct other near-horizon geometries of extremal black ring topology to test these ideas.  The general classification method presented in this paper should help towards this.  \\

\par \noindent \emph{Acknowledgements-}  HK is supported by NSERC and the Pacific Institute for the Mathematical Sciences. JL is supported by an EPSRC Career Acceleration Fellowship.

\appendix
\section{Near-horizon geometry of most general known extremal black string}
\label{kerrstring}
Consider the six-parameter, asymptotically $\mathbb{R}^{1,3} \times S^1$ black string solution presented in \cite{CBSV}. From the five-dimensional point of view, the string solution is parameterized by its mass, electric charge, magnetic charge, linear momentum along the string, angular momentum transverse to the string and the circumference of the string at infinity. The extremal limit is achieved by taking $|a|=m$, and we will assume $a > 0$.  After a tedious calculation one can construct the associated near-horizon geometry.  The resulting solution is parameterized by five parameters: $(a,R)$ which have dimensions of  length, and three dimensionless boost parameters $(\beta, \delta, \gamma)$.
The metric functions are given by
\begin{eqnarray} \label{expsola}
L^2 \gamma_{ij}(x) d\phi^i d\phi^j &=& \frac{\ell^2 (1-x^2)\Gamma(x)}{ P(x)}\left[u_1 d\phi^2-\frac{R s_\gamma d\phi^1}{a} \right]^2  \nonumber \\ && + \frac{a^6 P(x)}{\ell^4 \Gamma(x)^2}\left[(c_\gamma - s_\gamma \Omega(x))\frac{R d\phi^1}{a} + (u_0 + u_1 \Omega(x)) d\phi^2\right]^2
\end{eqnarray} where  $c_\beta = \cosh\beta$, $s_\beta = \sinh \beta$, etc. and
\begin{eqnarray}
&& \Gamma(x) = \frac{a^2}{\ell^2} [f_1(x) - (1-x^2)]  \\
&& P(x) = (f_4(x) - (1-x^2))(f_1(x) - (1-x^2)) - f_2(x)^2  \\
&& \Omega(x) = \frac{f_5(x)(f_1(x) - (1-x^2)) - f_2(x) f_3(x)}{P(x)}  \; .
 \end{eqnarray} It can be checked that
 \begin{equation}
 u_0 + u_1 \Omega(x) = (1-x^2)\frac{b(x)}{P(x)}
 \end{equation} for a quadratic function $b(x)$.
Note that the invariantly defined function $\Gamma$ is quadratic, consistent with our general findings.  As in our general analysis, the angles $\phi^i$ have period $2\pi$ and $-1 \leq x \leq 1$. The local metric smoothly extends to a cohomogeneity-one metric on $S^1 \times S^2$ and $\partial / \partial \phi_2$ vanishes at the poles of the $S^2$.
We also have defined the constant length scales
\begin{eqnarray}
\ell^2 &=& 2 a^2 c_\delta^3\left[ c_\gamma\left(c_\beta^4 + s_\beta^4 + \frac{6 c_\beta^2 s_\beta^2 s_\delta^2}{1 + 3c_\delta^2}\right) - \frac{ 2 c_\beta s_\beta (c_\beta^2+ s_\beta^2) s_\gamma}{\sqrt{ 1 + 3c_\delta^2}}\right]  \\
L^3 &=& R \ell^2
\end{eqnarray} and constants
\begin{eqnarray}
u_1 = 2 c_\delta^3  \left[(c_\beta^2 + s_\beta^2)^2 - \frac{8 c_\beta^2 s_\beta^2}{1 + 3 c_\delta^2} \right] \;, \qquad
u_0= -\frac{4 c_\beta s_\beta c_\delta^3 (c_\beta^2 + s_\beta^2)}{\sqrt{1 + 3c_\delta^2}}  \; .
\end{eqnarray}
The functions $f_i(x)$ are \emph{linear} in $x$ and given by
\begin{eqnarray}
f_1(x) &=& 2 c_\delta^2\left( \frac{2c_\beta^2 s_\beta^2 s_\delta^2}{1+3c_\delta^2} + c_\beta^4 + s_\beta^4 + \frac{2 c_\beta s_\beta s_\delta x}{\sqrt{1 + 3c_\delta^2}}\right) \\
f_2(x) &=& 2 c_\delta s_\delta (c_\beta^2 + s_\beta^2) \left(\frac{2 c_\delta^2 s_\beta c_\beta}{\sqrt{1+3c_\delta^2}}
 + s_\delta x \right)\\
f_3(x) &=& 2 c_\delta\left( \frac{2 s_\delta s_\beta^2c_\beta^2(1 + c_\delta^2)}{1 + 3c_\delta^2} - (c_\beta^4 + s_\beta^4)s_\delta + \frac{2 s_\beta c_\beta x}{\sqrt{1 + 3c_\delta^2}}\right) \\
f_4(x) &=& 2\left(c_\delta^4 + s_\delta^4 + 2c_\beta^2 s_\beta^2 (c_\delta^2 + s_\delta^2)^2  + \frac{2 s_\delta^2 s_\beta^2 c_\beta^2}{1 + 3c_\delta^2} - \frac{2 s_\delta s_\beta c_\beta x}{\sqrt{1 + 3c_\delta^2}} \right) \\
f_5(x) &=& 2(c_\beta^2 + s_\beta^2)\left( \frac{2 c_\delta^4 s_\beta c_\beta}{\sqrt{1 + 3c_\delta^2}} -  s_\delta^3  x \right)  \; .
\end{eqnarray}
The potentials associated with the field strength are
\begin{equation}
b_1(x) = \frac{a^2 R }{ L \ell^2 \Gamma(x)} \left[ c_\gamma f_2(x) - s_\gamma f_3(x) \right] \qquad
  b_2(x) =  \frac{a b_2^0}{L} + \frac{a^3}{L\ell^2 \Gamma(x)}\left[ u_0 f_2(x) + u_1 f_3(x)\right], \
\end{equation} and
\begin{equation}
\mu(x) =\frac{ a L }{\ell^2 \Gamma(x)}\left[-c_\delta u_0  + 2c_\delta^2 s_\delta (c_\beta^2 + s_\beta^2) x - \frac{b_2^0 a^2}{ \ell^2} \left(c_\gamma f_2(x) - s_\gamma f_3(x)\right)\right]
\end{equation} where we have defined the constant
\begin{equation}
b^0_2= 2  c_\delta^2 s_\delta\left(\frac{12 c_\beta^2 s_\beta^2 c_\delta^2}{1 + 3c_\delta^2}  + 1 \right) \; .
\end{equation}
The shifted twist potentials $\tilde{Y}_i = Y_i - \mu b_i$ take the simple form (setting $\gamma =0$ for convenience)
\begin{equation}
\tilde{Y}_1(x) = \frac{R a }{\ell^2 \Gamma(x)}\left(m_0 + m_1 x \right)
\end{equation} where
\begin{eqnarray}
m_1 = -\frac{4 c_\beta s_\beta c_\delta^3}{\sqrt{1+3c_\delta^2}} \;, \qquad m_0 =  \frac{8c_\beta^2 s_\beta^2 c_\delta^3 s_\delta}{1 + 3c_\delta^2}
\end{eqnarray} and
\begin{equation}
\tilde{Y}_2(x) = \frac{a u_0}{R} \tilde{Y}_1(x) + \frac{a^2}{\ell^2\Gamma(x) }(n_0 + n_1 x)
\end{equation} where
\begin{eqnarray} \label{expsolb}
n_1 &=& -\frac{4 (c_\beta^2 + s_\beta^2)c_\delta^3 ( a_1 - c_\delta a_2)^2}{(1 + 3c_\delta^2)( c_\delta a_1 - a_2)} \;, \qquad
n_0 = \frac{8 c_\beta s_\beta (c_\beta^ 2 + s_\beta^2) c_\delta^5 s_\delta(a_1^2 - a_2^2)}{ (1+3c_\delta^2)^{3/2} ( c_\delta a_1 - a_2)} \\
a_1 &=& 1 + 3c_\delta^2 + 12 c_\delta^2 c_\beta^2 s_\beta^2 \;, \qquad a_2 = 4 c_\delta c_\beta^2 s_\beta^2 \;.
\end{eqnarray} Given the scalars, it is straightforward to extract the remaining near-horizon data $(k^i,e)$ from (\ref{NHdata1}) and (\ref{NHdata2}):
\begin{equation}
k^{\phi_1} = \frac{2 L^2 a^3 m_1 m_0}{\ell^4 R s_\delta} \;, \qquad k^{\phi_2} = -\frac{2L^2 c_\delta^3(c_\beta^2 + s_\beta^2)}{\ell^2 u_1} + \frac{s_\gamma R}{a u_1}k^{\phi_1} \;
\end{equation}
\begin{equation}
e =  \frac{16a^3 L s_\beta^2 c_\beta^2 c_\delta^5 s_\delta}{\ell^4}\left[\frac{c_\gamma(c_\beta^2 + s_\beta^2)}{1 + 3c_\delta^2} + \frac{2s_\beta c_\beta s_\gamma}{({1 + 3 c_\delta^2})^{3/2}}\right] \; .
\end{equation}
From the near-horizon data above we can immediately compute the dipole charge using~(\ref{NHD}) taking into account~(\ref{rescalemin}) (recall $v=(0,1)$):
\begin{equation}
\mathcal{D} = \frac{4\sqrt{3} a s_\beta c_\beta c_\delta^2}{\sqrt{1+ 3c_\delta^2}} \; .
\end{equation} Similarly, using the gauge-invariant formula~(\ref{Q5d}) one can compute the electric charge
\begin{equation}\label{QBS}
Q = \frac{2\sqrt{3} \pi R a s_\delta c_\delta}{G}\left[(c_\beta^2 + s_\beta^2)c_\gamma + \frac{2s_\beta c_\beta  s_\gamma}{\sqrt{1 + 3c_\delta^2}} \right] \; .
\end{equation}
Note these charges agree with those of the full string given in~\cite{CBSV}. The condition for the black string solution to have vanishing tension is \cite{CBSV}
\begin{equation} \label{tensionzero}
\frac{2 s_\beta c_\beta s_\gamma c_\gamma}{\sqrt{1+3c_\delta^2}} + \frac{(c_\beta^2 + s_\beta^2)}{4}\left[c_\gamma^2 + 1 - 3 s_\gamma^2(c_\delta^2 + s_\delta^2)\right] = 0  \; .
\end{equation} It has been shown in~\cite{CBSV} that there are two values of $\gamma$ such that this condition is satisfied, with no further conditions on the remaining parameters $(a,R,\beta,\delta)$.

Note that requiring $Q=0$ implies that either $\delta=0$ or the factor in the square brackets in (\ref{QBS}) vanishes. It is straightforward to verify that only the former possibility is compatible with the tensionless condition~(\ref{tensionzero}).

Finally, we note that this family of near-horizon geometry solutions in fact contains the locally AdS$_3\times S^2$ near-horizon geometry. In the parameterisation used here, this arises as the limit $a\to 0$ with $a e^{2\beta}, \delta, \gamma, R$ fixed (so $\beta \to \infty$), see also~\cite{CBSV}. Notice this limit is well defined even for the $\delta=0$ solution.

\section{Fibre metric at axes of symmetry}
Consider the behaviour of $\gamma_{ij}$ as $x \to \pm 1$ which must be non-singular since it is an invariant of the solution $\gamma(m_i,m_j)$. We begin by obtaining expressions for the scalars $b^i$ as $x \to \pm 1$. From (\ref{bup}) we find
\begin{equation}
b^i =  \frac{b^i_{\pm}}{1-x^2}  + O(1)  \qquad  \textrm{as }  x \to \pm 1
\end{equation}
where $b^i_{\pm}$ are constants given by
\bea
 b^i_+ &=& \frac{2\sqrt{2} (S_{17} S_{23} -S_{13} S_{27})( S_{1\phantom{i}}^i S_{23} - S_{2 \phantom{i}}^i S_{13})}{S_{13}^2 + S_{23}^2 }\\
b^i_- &=& \frac{2\sqrt{2} (S_{37} S_{43} -S_{33} S_{47})( S_{3\phantom{i}}^i S_{43} - S_{4 \phantom{i}}^i S_{33})}{S_{33}^2 + S_{43}^2 }
\eea
where we have used (\ref{mupm}). This allows one to deduce an explicit expression for $b^ib_i$ and hence its behaviour near the endpoints
\be
b^ib_i=  \frac{b^i_{\pm} b^{\pm} _i}{1-x^2} +O(1) \qquad \textrm{as } x \to \pm 1
\ee
where the constants $b^i_{\pm} b^{\pm} _i$ simplify to
\bea
\label{bsqpm1}
b_i^+b^i_+ = \frac{2\sqrt{2}\eta_{ij}S_{1}^{\phantom{1}i} S_{2}^{\phantom{5}j} (S_{13}S_{27}-S_{17}S_{23})}{ S_{13}^2+S_{23}^2}  ,  \quad
b_i^-b^i_- = \frac{2\sqrt{2}\eta_{ij}S_{3}^{\phantom{1}i} S_{4}^{\phantom{5}j} (S_{33}S_{47}-S_{37}S_{43})}{ S_{33}^2+S_{43}^2}   \; .
\eea  The quantity $\gamma M_{IJ}$ is guaranteed to have a finite limit as $x\to \pm 1$ as can be seen from our explicit solution for $M$ (\ref{Msoln}). In fact we have (for all indices):
\be
 M_{IJ} =  \frac{m^{\pm}_{IJ}}{1-x^2} +O(1) \qquad \qquad  \qquad x \to \pm 1
\ee
where $m^{\pm}_{IJ}$ are constants given by
\be
m^+_{IJ}= 4(S_{1I }S_{1J}+S_{2I }S_{2J}) \qquad m^-_{IJ}= 4(S_{3I }S_{3J}+S_{4I}S_{4J})    \; .
\ee
Note that $\gamma M_{IJ} \to \alpha^2 m_{IJ}^{\pm} \Gamma_{\pm}^{-1}$ as $x \to \pm 1$.
From (\ref{gammaij}) we can now see that as $x\to \pm 1$
\bea
\label{Ndef}
(\gamma+\mu^2) \gamma_{ij}  &=&   {N}^ \pm_{ij}  + O[(1-x^2)]\\
{N}^ \pm_{ij}   &\equiv &\frac{\alpha^2 m^{\pm}_{ij}}{\Gamma_{\pm}} -\tilde{Y}^{\pm}_i \tilde{Y}_j^{\pm} -\frac{b^k_\pm b_k^{\pm}b_i^{\pm}b_j^{\pm}}{\Gamma_{\pm}}+\frac{2\mu_{\pm} b^{\pm}_{(i} \eta_{j)k}b_\pm^k}{\Gamma_\pm}
\eea
and $N^\pm_{ij}$ are finite (possibly vanishing) as required by regularity.  In fact one may derive a simple expression for ${N}^\pm_{ij}$. First note that using (\ref{Ypm}) one can show
\be
\label{iden1}
\frac{\alpha^2 m^\pm_{ij}}{\Gamma_\pm}-\tilde{Y}_i^\pm\tilde{Y}_j^\pm= \alpha^{\pm}_i \alpha^{\pm}_j
\ee
where we have defined
\be
\alpha^+_i \equiv  \frac{S_{23}S_{1i}-S_{13}S_{2i}}{S_{13}^2+S_{23}^2} \qquad \qquad  \alpha^-_i \equiv  \frac{S_{43}S_{3i}-S_{33}S_{4i}}{S_{33}^2+S_{43}^2}    \; .
\ee Now using the $i$7 component of (\ref{Mexplicit}) one finds
\begin{equation}
\label{relation1}
 b_i^\pm (b_k^\pm b^k_\pm) - \mu_\pm \eta_{ik}b^{k}_\pm  + \mu_\pm \tilde{Y}_i^\pm = \frac{m_{i7}^\pm}{\sqrt{2}} \; .
\end{equation} Multiplying this by $b^+_j$ and using (\ref{constraint1}) leads to, after some calculation:
\be
{N}^\pm_{ij} = \left[ \alpha^{\pm}_i + \left(\frac{b_k^\pm b^k_\pm}{\Gamma_\pm}\right)^{1/2} b^{\pm}_i \right] \left[ \alpha^{\pm}_j + \left(\frac{b_k^\pm b^k_\pm}{\Gamma_\pm}\right)^{1/2} b^{\pm}_j \right]
\ee
which is rank-1. This shows that at the endpoints, when $\mu \neq 0$, $\gamma_{ij}$ is indeed rank-1, as is required from our general analysis.  The $\mu=0$ case can be treated similarly but requires a higher order calculation.


\begin{thebibliography}{99}
{\small

\bibitem{SV}
  A.~Strominger, C.~Vafa,
  ``Microscopic origin of the Bekenstein-Hawking entropy,''
  Phys.\ Lett.\  {\bf B379 } (1996)  99-104.
  [hep-th/9601029].

 \bibitem{CM}
  C.~G.~Callan, J.~M.~Maldacena,
  ``D-brane approach to black hole quantum mechanics,''
  Nucl.\ Phys.\  {\bf B472 } (1996)  591-610.
  [hep-th/9602043].

\bibitem{BMPV}
  J.~C.~Breckenridge, R.~C.~Myers, A.~W.~Peet, C.~Vafa,
  ``D-branes and spinning black holes,''
  Phys.\ Lett.\  {\bf B391 } (1997)  93-98.
  [hep-th/9602065].

\bibitem{Her}
  C.~A.~R.~Herdeiro,
  ``Spinning deformations of the D1 - D5 system and a geometric resolution of closed timelike curves,''
  Nucl.\ Phys.\  {\bf B665 } (2003)  189-210.
  [hep-th/0212002].

\bibitem{MS1}
  J.~M.~Maldacena, A.~Strominger,
  ``Statistical entropy of four-dimensional extremal black holes,''
  Phys.\ Rev.\ Lett.\  {\bf 77 } (1996)  428-429.
  [hep-th/9603060].


\bibitem{ER}
  R.~Emparan and H.~S.~Reall,
  ``A rotating black ring in five dimensions,''
  Phys.\ Rev.\ Lett.\  {\bf 88} (2002) 101101
  [arXiv:hep-th/0110260].

\bibitem{emparan}
  R.~Emparan,
  ``Rotating circular strings, and infinite non-uniqueness of black rings,''
  JHEP {\bf 0403} (2004) 064
  [arXiv:hep-th/0402149].

  \bibitem{EEF}
  H.~Elvang, R.~Emparan and P.~Figueras,
  ``Non-supersymmetric black rings as thermally excited supertubes,''
  JHEP {\bf 0502} (2005) 031
  [arXiv:hep-th/0412130].

  \bibitem{EEMR}
  H.~Elvang, R.~Emparan, D.~Mateos and H.~S.~Reall,
  ``A supersymmetric black ring,''
  Phys.\ Rev.\ Lett.\  {\bf 93} (2004) 211302
  [arXiv:hep-th/0407065].

  \bibitem{CH}
  K.~Copsey and G.~T.~Horowitz,
  ``The role of dipole charges in black hole thermodynamics,''
  Phys.\ Rev.\  D {\bf 73} (2006) 024015
  [arXiv:hep-th/0505278].

\bibitem{Fig1}
  P.~Figueras, E.~Jamsin, J.~V.~Rocha, A.~Virmani,
  ``Integrability of Five Dimensional Minimal Supergravity and Charged Rotating Black Holes,''
  Class.\ Quant.\ Grav.\  {\bf 27 } (2010)  135011.
  [arXiv:0912.3199 [hep-th]].

    \bibitem{PS}
  A.~A.~Pomeransky and R.~A.~Sen'kov,
  ``Black ring with two angular momenta,''
  arXiv:hep-th/0612005.

  \bibitem{BSat}
  H.~Elvang, P.~Figueras,
  ``Black Saturn,''
  JHEP {\bf 0705}, 050 (2007).
  [hep-th/0701035].

\bibitem{HOT}
  K.~Hanaki, K.~Ohashi, Y.~Tachikawa,
  ``Comments on charges and near-horizon data of black rings,''
  JHEP {\bf 0712 } (2007)  057.
  [arXiv:0704.1819 [hep-th]].

\bibitem{LP}
  J.~Lewandowski, T.~Pawlowski,
  ``Extremal isolated horizons: A Local uniqueness theorem,''
  Class.\ Quant.\ Grav.\  {\bf 20 } (2003)  587-606.
  [gr-qc/0208032].

\bibitem{KL4d}
 H.~K.~Kunduri and J.~Lucietti,
 ``Uniqueness of near-horizon geometries of rotating extremal AdS(4) black
  holes,''
  Class.\ Quant.\ Grav.\  {\bf 26} (2009) 055019
  [arXiv:0812.1576 [hep-th]].

\bibitem{Reall}
  H.~S.~Reall,
  ``Higher dimensional black holes and supersymmetry,''
  Phys.\ Rev.\  {\bf D68 } (2003)  024024.
  [hep-th/0211290].

  \bibitem{KLR:2006}
  H.~K.~Kunduri, J.~Lucietti, H.~S.~Reall,
  ``Do supersymmetric anti-de Sitter black rings exist?,''
  JHEP {\bf 0702 } (2007)  026.
  [hep-th/0611351].

  \bibitem{KL:2007}
  H.~K.~Kunduri, J.~Lucietti,
  ``Near-horizon geometries of supersymmetric AdS(5) black holes,''
  JHEP {\bf 0712}, 015 (2007).
  [arXiv:0708.3695 [hep-th]].

\bibitem{GMR}
  J.~B.~Gutowski, D.~Martelli, H.~S.~Reall,
  ``All Supersymmetric solutions of minimal supergravity in six- dimensions,''
  Class.\ Quant.\ Grav.\  {\bf 20 } (2003)  5049-5078.
  [hep-th/0306235].

\bibitem{GP}
  J.~Gutowski, G.~Papadopoulos,
  ``Heterotic Black Horizons,''
  JHEP {\bf 1007 } (2010)  011.
  [arXiv:0912.3472 [hep-th]].

   \bibitem{KLvac}
  H.~K.~Kunduri and J.~Lucietti,
  ``A classification of near-horizon geometries of extremal vacuum black
  holes,''
  J.\ Math.\ Phys.\  {\bf 50} (2009) 082502
  [arXiv:0806.2051 [hep-th]].

  \bibitem{KLstatic}
  H.~K.~Kunduri and J.~Lucietti,
  ``Static near-horizon geometries in five dimensions,''
  Class.\ Quant.\ Grav.\  {\bf 26} (2009) 245010
  [arXiv:0907.0410 [hep-th]].

  \bibitem{HI}
  S.~Hollands and A.~Ishibashi,
  ``All vacuum near horizon geometries in arbitrary dimensions,''
  arXiv:0909.3462 [gr-qc].

 \bibitem{CJLP}
  E.~Cremmer, B.~Julia, H.~Lu, C.~N.~Pope,
  ``Higher dimensional origin of D = 3 coset symmetries,''
  [hep-th/9909099].

\bibitem{CJ:1979}
  E.~Cremmer, B.~Julia,
  ``The SO(8) Supergravity,''
  Nucl.\ Phys.\  {\bf B159}, 141 (1979).


  \bibitem{KLR}
  H.~K.~Kunduri, J.~Lucietti and H.~S.~Reall,
  ``Near-horizon symmetries of extremal black holes,''
  Class.\ Quant.\ Grav.\  {\bf 24} (2007) 4169
  [arXiv:0705.4214 [hep-th]].

\bibitem{Maison}
  D.~Maison,
  ``Ehlers-harrison Type Transformations For Jordan's Extended Theory Of Gravitation,''
  Gen.\ Rel.\ Grav.\  {\bf 10 } (1979)  717-723.

 \bibitem{Gauntlett}
  J.~P.~Gauntlett, J.~B.~Gutowski, C.~M.~Hull, S.~Pakis, H.~S.~Reall,
  ``All supersymmetric solutions of minimal supergravity in five- dimensions,''
  Class.\ Quant.\ Grav.\  {\bf 20 } (2003)  4587-4634.
  [hep-th/0209114].

\bibitem{TYI}
  S.~Tomizawa, Y.~Yasui and A.~Ishibashi,
  ``A uniqueness theorem for charged rotating black holes in five-dimensional
  minimal supergravity,''
  Phys.\ Rev.\  D {\bf 79} (2009) 124023
  [arXiv:0901.4724 [hep-th]].

  \bibitem{FKLR}
  P.~Figueras, H.~K.~Kunduri, J.~Lucietti and M.~Rangamani,
  ``Extremal vacuum black holes in higher dimensions,''
  Phys.\ Rev.\  D {\bf 78} (2008) 044042
  [arXiv:0803.2998 [hep-th]].

   \bibitem{CBSV}
  G.~Compere, S.~de Buyl, S.~Stotyn and A.~Virmani,
  ``A General Black String and its Microscopics,''
  arXiv:1006.5464 [hep-th].

 \bibitem{CBJV}
G.~Compere, S.~de Buyl, E.~Jamsin and A.~Virmani,
  ``G2 Dualities in D=5 Supergravity and Black Strings,''
  Class.\ Quant.\ Grav.\  {\bf 26} (2009) 125016
  [arXiv:0903.1645 [hep-th]].


  \bibitem{CWV}
  G.~Compere, W.~Song and A.~Virmani,
  ``Microscopics of Extremal Kerr from Spinning M5 Branes,''
  arXiv:1010.0685 [hep-th].


\bibitem{Larsen}
  F.~Larsen,
  ``Entropy of thermally excited black rings,''
  JHEP {\bf 0510} (2005) 100
  [arXiv:hep-th/0505152].

\bibitem{emparan2}
  R.~Emparan,
  ``Exact Microscopic Entropy of Non-Supersymmetric Extremal Black Rings,''
  Class.\ Quant.\ Grav.\  {\bf 25 } (2008)  175005.
  [arXiv:0803.1801 [hep-th]].

\bibitem{nil}
  S.~-S.~Kim, J.~L.~Hornlund, J.~Palmkvist, A.~Virmani,
  ``Extremal solutions of the S3 model and nilpotent orbits of G2(2),''
  JHEP {\bf 1008 } (2010)  072.
  [arXiv:1004.5242 [hep-th]].

\bibitem{CEvP}
  M.~M.~Caldarelli, R.~Emparan, B.~Van Pol,
  ``Higher-dimensional Rotating Charged Black Holes,''
  JHEP {\bf 1104 } (2011)  013.
  [arXiv:1012.4517 [hep-th]].


}
\end{thebibliography}
\end{document}